\documentclass[aps,prmater,numerical,reprint,superscriptaddress]{revtex4-2}
\usepackage[utf8]{inputenc}
\usepackage{amsmath}
\usepackage{amsfonts}
\usepackage{xcolor}
\usepackage{hyperref}
\usepackage{graphicx}% Include figure files
\usepackage{comment}
\usepackage{commath}
\usepackage{multirow}
\usepackage{natbib}
\usepackage[capitalize]{cleveref}
\usepackage{autonum}
\usepackage[normalem]{ulem}

\makeatletter
\autonum@generatePatchedReferenceCSL{Cref}
\makeatother

\crefname{section}{Sect.}{Sects.}

\begin{document}

\title{Analytic elastic constants in molecular calculations: \\ Finite strain, non-affine displacements, and many-body interatomic potentials}
% \titlerunning{Elastic constants}
\author{Jan~Grießer}
\affiliation{Department of Microsystems Engineering, University of Freiburg, Georges-K\"ohler-Allee 103, 79110 Freiburg, Germany}

\author{Lucas~Fr\'{e}rot}
\affiliation{Department of Microsystems Engineering, University of Freiburg, Georges-K\"ohler-Allee 103, 79110 Freiburg, Germany}

\author{Jonas~A.~Oldenstaedt}
\affiliation{Department of Microsystems Engineering, University of Freiburg, Georges-K\"ohler-Allee 103, 79110 Freiburg, Germany}

\author{Martin~H.~M\"user}
\affiliation{Department of Material Science and Engineering, Saarland University, Campus C6 3, 66123 Saarbrücken, Germany}

\author{Lars~Pastewka}
\email[Corresponding author: ]{lars.pastewka@imtek.uni-freiburg.de}
\affiliation{Department of Microsystems Engineering, University of Freiburg, Georges-K\"ohler-Allee 103, 79110 Freiburg, Germany}
\affiliation{Cluster of Excellence livMatS, Freiburg Center for Interactive Materials and Bioinspired Technologies, University of Freiburg, 79110 Freiburg, Germany}

% Use this for 3N-vectors
\let\vczech\v  % needed when \v is used in biblio to for Czech accents
\renewcommand{\v}{\mathbf}
% Use this for 3-vectors
\newcommand{\vv}{\vec}
% Use this for 3N x 3N matrices (tensors)
\renewcommand{\t}{\mathbb}
% Use this for 3 x 3 matrices (tensors)
\renewcommand{\tt}{\underline}
% Use this for 3 x 3 x 3 x 3 tensors
\newcommand{\ff}[1]{\underline{\underline{#1}}}
% Use this symbol for elastic constants without non-affine contribution 
\newcommand{\cc}{c}
% Use this symbol for elastic constants with non-affine displacements
\newcommand{\CC}{C}

% Lars Pastewka
\newcommand{\lp}[1]{\textcolor{red}{#1}}
% Jan Grießer
\newcommand{\jg}[1]{\textcolor{blue}{#1}}
% Martin Müser
\newcommand{\mm}[1]{\textcolor{green}{#1}}
% Lucas Frérot
\newcommand{\lf}[1]{\textcolor{orange}{#1}}
% Jonas Oldenstaedt
\newcommand{\jo}[1]{\textcolor{magenta}{#1}}
\date{\today}

\begin{abstract}
Elastic constants are among the most fundamental and important properties of solid materials, which is why they are routinely characterized in both experiments and simulations.
While conceptually simple, the treatment of elastic constants is complicated by two factors not yet having been concurrently discussed: finite-strain and non-affine, internal displacements.
Here, we revisit the theory behind zero-temperature, finite-strain elastic constants and extend it to explicitly consider non-affine displacements.
We further present analytical expressions for second-order derivatives of the potential energy for two-body and generic many-body interatomic potentials, such as cluster and empirical bond-order potentials.
Specifically, we revisit the elastic constants of silicon, silicon carbide and silicon dioxide under hydrostatic compression and dilatation.
Based on existing and new results, we outline the effect of multiaxial stress states as opposed to volumetric deformation on the limits of stability of their crystalline lattices.
\end{abstract}

\maketitle

\section{Introduction}

Elastic constants describe the stress needed to reversibly strain a solid.
If the solid is already strained, they correspond to the extra stress needed to deform it further.
They are among the most fundamental and important properties of solids and are hence routinely computed in atomic-scale calculations. 
Unfortunately, a rigorous computation of these deceivingly simple properties is complicated by two factors.
First, neither stress nor strain are uniquely defined, except in the immediate vicinity of zero external stress, so that numerous definitions of elastic tensors exist.
Which of the various definitions of finite-strain elastic constants matters depends on the property of interest.
For example, the elastic tensors used to deduce the generally polarization- and direction-dependent sound velocity~\citep{huang_atomic_elasitcity_1950,barron_second-order_1965,thurston_wave_1965,wallace_thermoelasticity_1967} are different from those used for the analysis of lattice stability~\citep{born_thermodynamics_1939,born_stability1_1940,born_dynamical_1954,wallace_thermoelasticity_1967,wallace_thermoelasticity_1967, wallace_thermoelastic_1970,hill_elasticity_and_stability_1975,hill_principles_1977,wang_crystal_1993,wang_mechanical_1995,grimvall_lattice_instabilities_2012}, which, according to Born, necessitates the elastic tensor to be positive definite. 
Second, atomic positions undergo non-affine internal relaxations in response to a macroscopic shape change, except in the case of highly symmetric crystals for which all Wykoff positions are fully determined by symmetry even in the deformed state. 
An extreme case for solids with non-affine internal relaxations are amorphous materials for which no single atomic position can be deduced from symmetry.

While the relevance of the various elastic tensors as well as the relations between them are well established, we find the treatments that we are aware of anything but transparent and complete. 
For example, we are not aware of any prior work clearly illustrating the existence and uniqueness of the generalization of the enthalpy or Gibb's free energy when a non-isotropic external stress is applied.
Moreover, the computation of non-affine displacements is computationally expensive for large systems unless analytic second-order derivatives of the potential are known---but pertinent expressions appear to be documented in the literature only for pair potentials. 
Consequently, elastic tensors of large disordered systems have so far been determined mainly when they could be modeled within the pair-potential approximation. 

The purpose of this paper is to develop finite-strain elasticity in the presence of external stresses and non-affine displacements.
The path towards this goal is as follows:
We first revisit the derivation of the Born stability criteria, then
we generalize the determination of non-affine displacements to finite strains using a variational formulation that is inspired by (continuum) stochastic homogenization techniques. 
Finally, we present explicit expressions for second-order derivatives of the potential energy for generic many-body potentials, in particular empirical bond-order potentials and cluster potentials.
We use those to discuss the lattice stability of silicon, silicon carbide, and silicon dioxide crystals under hydrostatic and multiaxial external stress.

\section{Finite deformation and thermodynamic potentials}
\subsection{Finite strain}
In this section, we assume a periodic simulation cell in $D$ dimensions, which can also serve as a representative volume element (RVE).
Its geometry is described by the h-matrix~\citep{parrinello_polymorphic_1981}, $\tt{h} \equiv (\vv{a}_1,\ldots,\vv{a}_D)$, whose $D$ columns $\vv{a}_i$ are $D$-dimensional vectors spanning the simulation cell. 
The volume of the simulation cell is then given by $V = \det(\tt{h})$.
Positions of atoms in the simulation cell can be stated in terms of either true coordinates, $\vv{r}_i$, or unitless, scaled coordinates, $\vv{s}_i$.
True and scaled coordinates of the $i = 1,\ldots,N$ atoms in the simulation cell are related through the equation $\vv{r}_i = \tt{h}\cdot\vv{s}_i$.

Let us define one particular h-matrix as the reference h-matrix and indicate its reference status with a circle as $\mathring{\tt{h}}$.
Similarly, we assume reference scaled atomic coordinates, $\mathring{\vv{s}}_i$.
The effect of an \emph{affine} deformation of the simulation cell can be described by the \emph{deformation gradient} $\tt{F}$ acting on the reference cell according to $\tt{h} = \tt{F}\cdot\mathring{\tt{h}}$ while leaving the $\mathring{\vv{s}}_i$ unchanged, $\vv{s}_i\equiv\mathring{\vv{s}}_i$.
In the affinely deformed cell, atomic coordinates thus read
\begin{equation}
    \tt{r}_i  =  \tt{F} \cdot  \mathring{\vv{r}}_i
    \qquad\text{or}\qquad
    r_{i\alpha} = F_{\alpha\beta} \mathring{r}_{i\beta},
    \label{eq:affine}
\end{equation}
where $\mathring{\vv{r}}_i = \mathring{\tt{h}}\cdot\mathring{\vv{s}}_i$.
We have written \cref{eq:affine} in tensor and index notation and use Einstein summation convention for repeated indices in the latter.
Atoms are identified by Roman indices, directions by Greek indices.
In tensor notation, there are no Greek indices but an arrow indicates a $D$-vector and an underline a $D\times D$ matrix.
We will switch between tensor and index notation throughout this article, using the notation that we deem appropriate for each equation.

In the absence of external (magnetic, electric, or strong gravitational) fields,
the potential energy is translationally and rotationally invariant, a property that 
is called \emph{objectivity} in the field of mechanics. 
The total potential energy of a system can then only depend on the set of distances ${r}_{ij}= \vert \vv{r}_i-\vv{r}_j \vert$ rather than on the absolute positions. This is also true for potentials beyond the pair-potential approximation, since knowledge of all $r_{ij}$ defines the system up to translation, rotation, and handedness.

It is important to note that absolute distances transform as $R_{ij} \equiv r_{ij}^2 = \mathring{\vv{r}}_{ij}\cdot \tt{F}^T \cdot \tt{F} \cdot \mathring{\vv{r}}_{ij}$, with the \emph{right Cauchy--Green deformation tensor} $\tt{F}^T \cdot \tt{F}$. 
The \emph{strain} in the system is often characterized by how much this deformation tensor deviates from the identity matrix, $\tt{1}$, giving the \emph{Green--Lagrange strain tensor} \cite{lai_continuum_mechanics_2009}
\begin{equation}
    \tt{\eta} \equiv \frac{1}{2} \left(\tt{F}^T \cdot \tt{F} - \tt{1}\right).
    \label{eq:greenlagrange}
\end{equation}
The Green--Lagrange strain is sometimes denoted by the symbol $\tt{E}$ in the solid mechanics literature.

We can write the deformation gradient in terms of the h-matrix as
\begin{equation}
    \tt{F} = \tt{h}\cdot\mathring{\tt{h}}^{-1}.
    \label{eq:deformation_gradient}
\end{equation}
The Green--Lagrange strain tensor then becomes
\begin{equation}
    \tt{\eta} = \frac{1}{2} \left(\mathring{\tt{h}}^{-T} \cdot \tt{h}^T \cdot \tt{h}\cdot\mathring{\tt{h}}^{-1} - \tt{1}\right).
    \label{eq:greenlagrangeh}
\end{equation}
The Green--Lagrange strain tensor is hence \emph{always} defined with respect to a reference h-matrix.
Moreover, the Green--Lagrange strain tensor is independent of rotation, that is, $\tt{\eta}$ remains unchanged upon rotation of the simulation cell.
This can be seen when rotating each vector spanning the $\tt{h}$ matrix in \cref{eq:greenlagrangeh} through the operation $\tt{R}\cdot\tt{h}$ , where $\tt{R}$ is a rotation matrix fulfilling $\tt{R}^T \cdot \tt{R} = \tt{1}$.  

Another measure for the deformation is the small--strain tensor \cite{lai_continuum_mechanics_2009}
\begin{equation}
\tt{\varepsilon} \equiv \frac{1}{2} (\tt{F} + \tt{F}^T) - \tt{1},
\label{eq:epsilon-to-eta}
\end{equation}
which does not remain unchanged under a rotation of the h-matrix.
Finally, we note that 
\begin{equation}
    \eta_{\alpha\beta} 
    = 
    \varepsilon_{\alpha\beta} + \frac{1}{2} \varepsilon_{\alpha\gamma}\varepsilon_{\gamma\beta},
    \label{eq:relation_eta_eps}
\end{equation}
from which the partial derivative of $\eta_{\alpha\beta}$ with respect to $\varepsilon_{\mu\nu}$ can be obtained as
\begin{equation}
    \frac{\partial \eta_{\alpha\beta}}{\partial \varepsilon_{\mu\nu}} = 
    \delta_{\alpha\mu} \delta_{\beta\nu} + \frac{1}{2}\left(\delta_{\alpha\mu} \varepsilon_{\nu\beta} + \varepsilon_{\alpha\mu} \delta_{\beta\nu}
    \right).
\end{equation}
Thus, for a second-order derivative of an arbitrary function $f$, it follows that
\begin{equation} \label{eq:diffRuleEulerLag}
    \left. \frac{\partial^2 f}{\partial\varepsilon_{\alpha\beta} \partial\varepsilon_{\mu\nu}} \right|_{\tt{\varepsilon}=0} =
    \left.\frac{\partial^2 f}{\partial\eta_{\alpha\beta} \partial\eta_{\mu\nu}}\right|_{\tt{\eta}=0} +
    \left.\frac{\partial f}{\partial \eta_{\alpha\mu}}\right|_{\tt{\eta}=0}\,\delta_{\beta\nu}.
    \label{eq:relation_eta_eps_second_deriv}
\end{equation}

\subsection{\label{sec:fsEnergy}Internal and free energy}
Assuming the initial structure to be fully relaxed mechanically, we define a zero-temperature energy
\begin{equation}
\label{eq:defineU}
    U(N,\tt{h}) = \min_{\{\vv{s}_i\}} u(\{\vv{s}_i\},\tt{h}).
\end{equation}
where $U(N,\tt{h})$ is the internal energy.
Here, the minimization is meant to implicitly contain the instruction ``next available minimum'' along an adiabatic change of the h-matrix from $\mathring{\tt{h}}$ to $\tt{h}$.
Unlike for a global minimum, this means that $U(N,\tt{h})$ depends on the atomic coordinates in the initial reference configuration and on the path from $\mathring{\tt{h}}$ to $\tt{h}$.

In a thermal (equilibrium) system, the instantaneous energy has to be replaced with the Helmholtz free energy $A$, which would be the proper thermodynamic potential that the internal coordinates would minimize in thermal equilibrium at fixed h-matrix. 
For a classical, single-atomic system, it is formally defined as
\begin{equation}
    \label{eq:freeEnergy}
    A(N,T,\tt{h}) = -k_\mathrm{B} T 
    \ln
    Z(N,T,\tt{h})
\end{equation}
with
\begin{equation}
    \label{eq:partitionFunction}
    Z(N,T,\tt{h}) 
    =
    \frac{1}{N!}
    \frac{1}{ \lambda_\textrm{B}^{3N}}\int \mathrm{d}\Gamma_\textrm{c} e^{-u(\Gamma_\textrm{c},\underline{h})/(k_\textrm{B}T)}.
\end{equation}
General phase-space averages of an observable $O$ are given by
\begin{equation}
    \label{eq:observable}
    \langle O \rangle = Z^{-1} \int \dif \Gamma_c O(\Gamma_c, \tt{h}) e^{-u(\Gamma_\textrm{c},\underline{h})/(k_\textrm{B}T)}.
\end{equation}
Here $k_\textrm{B}T$ is the thermal energy and $\lambda_\mathrm{B}$ the De Broglie wavelength, while $\Gamma_\textrm{c}$ represents the configurational phase space containing all Cartesian coordinates of the $N$ atoms, which are confined to the volume spanned by the h-matrix. 

When defining an elastic free energy of solids, it is generally necessary to constrain the phase-space integral in \cref{eq:partitionFunction,eq:observable}, since otherwise even macroscopic crystals would lose their ability to withstand non-isotropic stress and consequently fill any volume in a similar fashion as liquids do.
This is because in true equilibrium, solids can release static stress over long time scales, e.g., through the motion of lattice defects like vacancies or dislocations. 
Thus, in the above definition of the free energy it is necessary to keep in mind the concept of the separation of time scales and to  restrict the integration over phase space to only those configurations that can be reached from the initial or reference configuration without having to pass over (macroscopically) large energy barriers, as those configurations are not reached by thermal fluctuations on an experimental time scale.
Alternatively, one could conduct the phase-space integration within the harmonic approximation, or restrict the integral to the configurations having fixed bonding topography, or to basins in which the Hessian of the energy w.r.t. (scaled) atomic coordinates is positive definite, etc.
This then necessitates the introduction of additional collective variables describing the current local minimum for which to compute the free energy.
Ultimately, a certain ambiguity for the definition of a meaningful restricted free energy remains.
However, the supposedly justified hope is that final results are not critically affected by the choice of the restriction.

To make the definition of an elastic free energy at finite temperature more concrete, we constrain mean atomic positions of individual atoms, $\langle\vv{s}_i\rangle$ to $\vv{s}_i^\text{c}$~\cite{Ottinger2005-mq}, and choose $\vv{s}_i^\text{c}$ such that the constraint free energy, $a(\{\vv{s}_i^\text{c}\}; \tt{h})$, is minimized.
In a purely harmonic solid, the $\vv{s}_i^\text{c}$ are the equilibrium positions at zero temperature, but not generally in anharmonic solids.
At zero temperature, this free energy reduces to the internal energy.
In the following we drop the explicit superscript c from the equations, and implicitly refer to the constraint variables when writing $\vv{s}_i$ or $\vv{r}_i$.
To conclude, we assume not only the internal but also the free energy to depend on some (initial) reference configuration and provide definitions for any arbitrary h-matrix, including those where a given h-matrix is a strained configuration. 

\subsection{Stresses as conjugates to strain variables}
Stresses can be defined as the first-order derivative of the free energy with respect to a strain measure.
This leaves different stress tensors to be defined, because different strain tensors exist.
Formally, the stress for an arbitrary structure is given by
\begin{equation}
    \tt{\sigma}^{\eta} 
    \equiv \frac{1}{V}
    \frac{\partial A\left(\{\vv{s}_i\},\tt{h}\right)}{\partial \tt{\eta}},
    \label{eq:definition_stress}
\end{equation}
where the derivative is to be interpreted component-wise (i.e. it is the gradient).
Internal relaxation of the atomic coordinates during the small perturbation of the cell necessary to compute the derivative do not matter if the $\vv{s}_i$ minimize the free energy for the cell $\tt{h}$, which we will assume in the following.
Relaxation only matters at second order and will affect the elastic constants discussed below.
The nomenclature that was implicitly introduced uses upper indices to indicate the definition of the stress, i.e., the strain tensor with respect to which the first-order derivative of the free energy was taken. 
At this point, we abstain from associating pertinent expressions with the commonly used names, which are Kirchoff, nominal, first and second Piola--Kirchoff, and Biot stress~\cite{lai_continuum_mechanics_2009}.

The stresses obtained from \cref{eq:definition_stress} differ for different strain measures.
For example, using \cref{eq:epsilon-to-eta} we can derive the relation between the stresses obtained by using small--strain tensor and Green--Lagrange strain tensor as  
\begin{equation}
   \sigma^{\varepsilon}_{\alpha\beta} 
   =
   \frac{1}{V} 
   \frac{\partial  A}{\partial \eta_{\gamma\delta}}
   \frac{\partial \eta_{\gamma\delta}}{\partial \varepsilon_{\alpha\beta}}
   =
   \sigma^{\eta}_{\alpha\beta} + \frac{1}{2}\left(
   \sigma^{\eta}_{\alpha\gamma} \varepsilon_{\gamma\beta} + \varepsilon_{\alpha\gamma}
   \sigma^{\eta}_{\gamma\beta} 
   \right) + O(\varepsilon^2).
   \label{eq:PK2-to-Cauchy}
\end{equation}
Thus, $\tt{\sigma}^{\eta}$ is only identical to $\tt{\sigma}^{\varepsilon}$ if the stress is evaluated at zero strain, however, for an arbitrary $\mathring{\underline{h}}$. 
The resulting stress is called the \emph{Cauchy stress}, often also referred to as the true stress.

\subsection{\label{sec:thermodynamic-potentials}Generalized Gibbs free energy}

In thermodynamics, coupling a subsystem to an external bath is commonly done such that at least one extrinsic thermodynamic variable associated with the subsystem, e.g., the particle number $N$, is kept constant and at least one is relaxed, such as volume $V$, while its conjugate variable, in this case the hydrostatic pressure $p$, is constrained to a fixed value.
In this specific example, the \emph{extended} Gibbs free energy,
\begin{equation}
    g(N,p,T; V) = A(N,V,T) + pV.
    \label{eq:extended-gibbs-free-energy}
\end{equation}
would then be minimized with respect to the volume $V$ yielding the Gibbs free energy,
\begin{equation}
    \label{eq:gibbs-free-energy}
    G(N,p,T) = \min_{V} g(N,p,T; V).
\end{equation}
The value of $V$ minimizing $g(N,p,T; V)$, $V^\textrm{eq}$, is the expectation or equilibrium value of the volume. 
Note that  \cref{eq:gibbs-free-energy} has formally the form of a Legendre transformation.
Rather than stating $V^\textrm{eq}$ as a function of $p$, it is common to express pressure as a function of $V^\textrm{eq}$, i.e.,
\begin{equation}
    p \equiv p(V^\textrm{eq}) = -\left.\frac{\partial A}{\partial V}\right|_{V^\text{eq}},
\end{equation}
which follows from the requirement that $V$ minimizes the extended Gibb's free energy in \cref{eq:extended-gibbs-free-energy} at $V = V^\textrm{eq}$.

A physical realization of the just described situation could be a crystal---for the sake of simplicity described within the harmonic approximation to avoid discussions related to broken ergodicity or separation of time scales---embedded into an ideal gas, which acts as an infinitely large bath while exerting the pressure $p$ on the crystal. 
The term $W = pV$ is the work done on the gas to make room for the crystal, or, alternatively, the work done on the periodic boundary conditions imposed in a molecular simulation.
In the following, we will not distinguish between these two cases so that the h-matrix can be said to span our subsystem, which thus is implicitly assumed to be a parallelepiped.  

The gas can freely flow around our crystal, thus imposing purely hydrostatic conditions.
Unfortunately, it is not possible to construct a related expression for the work imposing a constant non-isotropic stress (which cannot be realized by imaging an ideal gas as a embedding medium) that would be independent of the current h-matrix. 
A general path-independent work has to be a function of the h-matrix so that the most general extended Gibbs free energy reads
\begin{equation}
    g(\underline{h}) = A(\underline{h}{) + W(\underline{h})},
    \label{eq:gen-gibbs}
\end{equation}
with the hydrodynamic case leading to the special choice of $W(\tt{h}) = p\,\det(\underline{h})$.
Before proceeding, we note that $g(\underline{h})$ can depend on many more variables than just $N,T$, which were dropped in the discussion.

One (formally) possible dependence of the work on $\underline{h}$ would be
\begin{equation}
\label{eq:workExample1} 
    W = 
    -\Pi_{11} h_{11} \mathring{h}_{22} \mathring{h}_{33} 
    -\Pi_{22} \mathring{h}_{11} h_{22} \mathring{h}_{33}
    - \Pi_{33} \mathring{h}_{11} \mathring{h}_{22} h_{33}.
\end{equation}
and the parameters $\Pi_{\alpha\beta}$ could be associated with the coefficients of a target stress tensor.
Minimizing \cref{eq:gen-gibbs} yields the equilibrium h-matrix $\tt{h}^\text{eq}$.
For $\mathring{\underline{h}}=\underline{h}^\textrm{eq}$,  $\underline{\Pi}$ can be associated with the externally imposed stress, which does not have to be isotropic.
Thus, in order to impose the three desired independent stress-tensor eigenvalues of a symmetric stress tensor  in a molecular simulation using \cref{eq:workExample1}, it is necessary to know ahead of time, or, alternatively, to identify by recursion the equilibrium h-matrix~\cite{miller_molecular_2016}.
Any deviation of $\mathring{\underline{h}}$ from $\underline{h}^\textrm{eq}$ will generally make the actual stress acting on the subsystem deviate from the target stress.
Note, however, that we can always interpret $\tt{\Pi}$ as the stress acting in some reference configuration with h-matrix $\tt{h}$.
This type of ``stress'' is commonly called the second Piola--Kirchhoff stress, as opposed to the ``true'' or Cauchy stress that acts in the equilibrium configuration.

It can be helpful to realize that imposing a constant, isotropic stress using 
\cref{eq:workExample1} will lead to different thermal fluctuations of the h-matrix---and thus to different stability conditions---than if the isotropic stress were imposed through $W = p\,\det{\underline{h}}$.
Both are determined by a second-order expansion of $g$ into h-matrix elements.
It may be concluded that individual stress tensor elements cannot be used as independent, intrinsic thermodynamic variables.
The stress acting on the surface of a subsystem embedded into an elastic medium (the ``bath''), say, an ideal, heterogeneous but anisotropic linearly elastic manifold, is not necessarily identical to that associated with a spatial average over the embedding medium.
Equilibrium merely requires $\nabla\cdot \tt{\sigma}$ to vanish.
In other words, the (equilibrium) bath stress is not necessarily the one acting on the surface of the subsystem. 
Even worse, in a real subsystem / bath system, the subsystem would generally not  maintain the shape of a parallelepiped, due to stress singularities near the edges and corners if subsystem and embedding differ in their elastic properties.
Any constant stress ensemble must therefore make simplifying assumption and rely on a hypothetical stress boundary condition, which is unlikely representative of the true situation.

To connect back to solid mechanics, we rewrite the thermodynamic potentials in terms of a strain and a reference h-matrix, $\mathring{\tt{h}}$.
Working with strain has the advantage that $W$ can be made objective.
This eliminates degeneracy in the thermodynamic potentials, but at the cost of introducing some (artificial) reference $\mathring{\tt{h}}$, which has no specific physical meaning unless chosen at zero stress.
A linear expansion of $W$ in $\eta$ yields
\begin{equation}
\label{eq:workExample1-finite-strain} 
    \begin{split}
        W = 
        -
        \mathring{V}
        \left(
        \Pi_{11} \eta_{11} +
        \Pi_{22} \eta_{22} + 
        \Pi_{33} \eta_{33}\right)
    \end{split}
\end{equation}
with $\mathring{V}=\det \mathring{\tt{h}}$, but this expression only corresponds to \cref{eq:workExample1} to linear order.

\subsection{Work on a hypothetical embedding medium}

In order to generalize the above discussion to non-orthorhombic deformation, described by some parameterized configuration path $\tt{h}(s)$, or, $\tt{\eta}(s)$, we adopt a continuum view on our RVE.
Specifically, our goal is to homogenize the response of the molecular RVE to that of an equivalent homogeneous elastic continuum.
This implies that the deformation throughout the \emph{homogenized} RVE is affine.
(For this discussion, it does not matter that the deformation in our reference molecular RVE, as discussed below, will be non-affine.)

We now compute the work $W$ performed on the embedding medium along some parameterized path $s$ at hypothetical constant external stress $\tt{\tau}$.
Note that $s$ can be interpreted as a form of time, in which case derivatives with respect to $s$ becomes rates and we will denote such derivatives with a dot, $\dot W=\dif W/\dif s$. 
As already discussed above, we call this embedding medium ``hypothetical'' because we ignore boundary effects that may arise from a mismatch of elastic properties between embedding medium and RVE.
We also ignore stress gradients that may be present in the embedding medium, depending on the macroscopic (at infinity) boundary conditions.
This hypothetical embedding medium can be imagined as consisting of infinitely repeated images of the RVE under perfectly matched boundary conditions.
Assuming that the current configuration of the RVE has h-matrix $\tt{h}(s)$, the work $W$ performed on the embedding medium along an arbitrary path $s$ is given by 
\begin{equation}
    W 
    =
    \int \dif W
    = 
    \int
    \dif s\,
    \dot{W}
    \label{eq:workintegral}
\end{equation}
where $\dot{W}$ is the power that is obtained by integrating the external forces $\dif \vv{f}(\vv{r})=-\tt{\tau}\cdot\hat{n}(\vv{r})\dif^2r$ times the velocity $\dot{\vv{u}}$ over the surface $\partial \tt{h}$ of our representative volume element.
In order to simplify our notation we omit the explicit dependency on $\vv{r}$ for the remaining part of this section.
The rate of work $\dot{W}$ is 
\begin{equation}
    \dot{W}
    =
    \int \dif \vv{f} \cdot \dot{\vv{u}}
    =
    -
    \int_{\partial \tt{h}(s)} \dif^2 r\, \dot{\vv{u}}\cdot\tt{\tau}\cdot\hat{n},
    \label{eq:work1}
\end{equation}
where $\hat{n}$ is the normal vector of the surface.
The tensor $\tt{\tau}$ is symmetric because of conservation of angular momentum.

Let us now assume that we have a Cauchy stress field $\tt{\sigma}$ throughout the RVE.
On the boundary $\tt{\sigma}=\tt{\tau}$.
We remain at mechanical equilibrium where $\nabla\cdot\tt{\sigma}=0$ along the deformation path.
To relate $\tt{\sigma}$ to the rate of work of external forces, we use the divergence theorem,
\begin{equation}
    \dot{W}
    =
    -
    \int_{\partial \tt{h}(s)}\dif^2 r\,
    \dot{\vv{u}}\cdot\tt{\sigma}\cdot\hat{n}
    =
    -
    \int_{\tt{h}(s)} \dif^3 r\,
    \nabla\cdot\left[\tt{\sigma}\cdot\dot{\vv{u}}\right].
\end{equation}
Under mechanical equilibrium this yields
\begin{equation}
    \dot{W} 
    =
    -\int_{\tt{h}(s)} \dif^3 r\,
    \tt{\sigma}:\nabla\dot{\vv{u}}
    =
    -\int_{\tt{h}(s)} \dif^3 r\,
    \tt{\sigma}:\dot{\tt{\varepsilon}},
    \label{eq:work2}
\end{equation}
where $\dot{\tt{\varepsilon}}$ is the strain-rate tensor.
The colon indicates a double contraction, that is,  $\tt{A}:\tt{B}=A_{\alpha\beta}B_{\beta\alpha}$.
The integral in \cref{eq:work2} is over the current configuration.

The tensor $\dot{\varepsilon}$ is defined with respect to the current h-matrix $\tt{h}(s)$ along the deformation path.
We cannot directly evaluate the integral \cref{eq:workintegral}, because the h-matrix (that enters the integration bounds) depends on $s$ in \cref{eq:work2}.
By introducing a reference h-matrix $\mathring{\tt{h}}$ and reference coordinates $\mathring{\vv{r}}$, we can express the velocity by $\dot{\vv{u}}(s)=\dot{\tt{F}}(s)\cdot\mathring{\vv{r}}=\dot{\tt{F}}(s) \cdot \tt{F}^{-1}(s)\cdot \vv{r}(s)$, where $\dot{\tt{F}}$ is the rate of change of the deformation gradient $\tt{F}$.
This yields
\begin{equation}
    \dot{\tt{\varepsilon}}
    =
    \frac{1}{2} \left[\frac{\partial \dot{\vv{u}}}{\partial \vv{r}} + \left(\frac{\partial \dot{\vv{u}}}{\partial \vv{r}} \right)^T\right]
    =
    \frac{1}{2} \left( \dot{\tt{F}} \cdot \tt{F}^{-1} + \tt{F}^{-T} \cdot \dot{\tt{F}}^T \right).
\end{equation}
From \cref{eq:greenlagrange} we get
\begin{equation}
    \dot{\tt{\eta}}
    =
    \frac{1}{2} \left( \dot{\tt{F}}^T \cdot \tt{F} + \tt{F}^{T} \cdot \dot{\tt{F}} \right),
\end{equation}
which we can use to write the power as a first order expression in $s$,
\begin{equation}
    \dot{W}
    =
    -\int_{\mathring{\tt{h}}} \dif^3\mathring{r}\, 
    \tt{\Pi}:\dot{\tt{\eta}},
    \label{eq:workincrcauchy}
\end{equation}
where now the integral is over our (arbitrarily chosen) reference cell with h-matrix $\mathring{\tt{h}}$.
Note that we have identified,
\begin{equation}
    \tt{\Pi}
    \equiv 
    J \tt{F}^{-1} \cdot \tt{\sigma} \cdot \tt{F}^{-T}
    \label{eq:pk2cauchy}
\end{equation}
with $J=\det\tt{F}$.
\Cref{eq:pk2cauchy}, which is often referred to as a pull-back operation~\citep{biot_mechanics_1965,lai_continuum_mechanics_2009,tadmor_modeling_2011,ted_belytschko_nonlinear_2014}, transforms the true (or Cauchy) stress $\tt{\sigma}$ into the second Piola--Kirchhoff stress $\tt{\Pi}$.
Identical expressions have been derived in the context of lattice stability~\citep{wang_crystal_1993,wang_mechanical_1995} and constant pressure molecular dynamics simulations~\citep{miller_molecular_2016}.

\subsection{Finite-strain elastic constants}

\begin{figure}
    \includegraphics[width=\columnwidth]{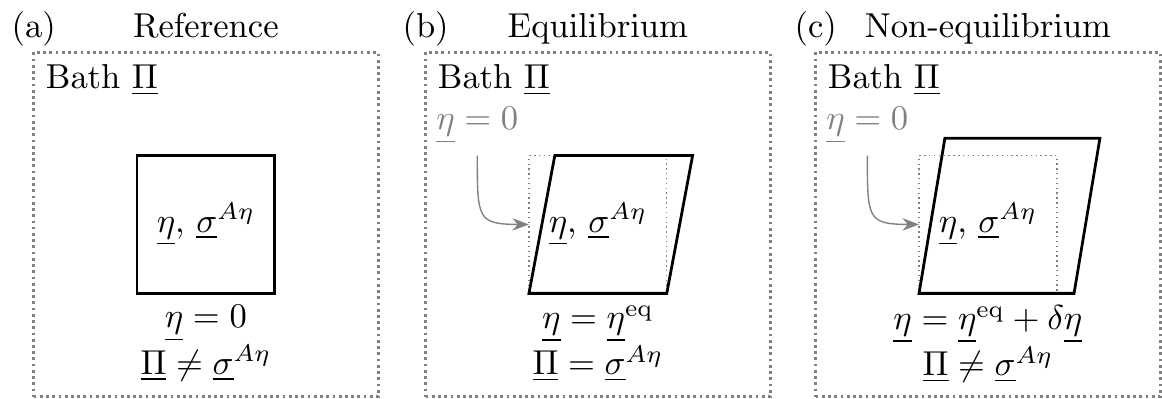}
	\caption{\label{fig:thermodynamics} Thermodynamics of elastic deformation. The external stress $\tt{\Pi}$ is a property of a hypothetical ideal embedding medium.
	Equilibrium is the case where embedding medium and system pressure are  equal, $\tt{\Pi}=\tt{\sigma}^{\eta}$. (a) Our reference configuration is not necessarily at equilibrium. (b) Equilibrium is obtained at some finite strain $\eta^\text{eq}$ where bath and system pressure are equal. (c) We make small deformations around this equilibrium condition to probe the elastic properties of the system at constant stress $\tt{\Pi}$.
 }
\end{figure}

We now assume that the RVE can be described by an equivalent homogeneous elastic continuum, which can only deform affinely.
Then $\dot{\tt{\eta}}$ is constant throughout the RVE and equivalent to the applied macroscopic strain.
This means we can replace $\tt{\Pi}(\vv{r})$ in \cref{eq:workincrcauchy} by its spatial average $\tt{\Pi}$.
Integration yields
\begin{equation}
    W = -\int_{s_0}^{s_1} \dif s \int_{\mathring{\tt{h}}} \dif^3\mathring{r}\,\tt{\Pi}:\dot{\tt{\eta}} = -\mathring{V} \tt{\Pi}:\tt{\eta},
\end{equation}
where $\eta$ is the total strain along path $s_0\to s_1$.
This gives our ad-hoc \cref{eq:workExample1-finite-strain} in the orthorhombic case.
The Gibbs free energy, in analogy to \cref{eq:gibbs-free-energy}, for arbitrary cell deformation is then
\begin{equation}
\begin{split}
    G(\tt{\Pi}; \mathring{\tt{h}}) 
    &=
    \min_{\tt{\eta}} g(\tt{\Pi}; \mathring{\tt{h}}; \tt{\eta}) \\
    g(\tt{\Pi}; \mathring{\tt{h}}; \tt{\eta}) 
    &=
    A(\mathring{\tt{h}}; \tt{\eta}) - \mathring{V} \tt{\Pi}:\tt{\eta},
    \label{eq:gibbs-finite-strain}
\end{split}
\end{equation}
where $\tt{\eta}=0$ is some reference configuration with h-matrix $h$ (see \cref{fig:thermodynamics}a), which is not necessarily in mechanical equilibrium (see \cref{fig:thermodynamics}b).
The quantity $g(\tt{\Pi};\mathring{\tt{h}};\tt{\eta})$ is the extended Gibbs free energy of a (small) system at strain $\tt{\eta}$.

We now strain the system against the constant external stress $\tt{\Pi}$ of a surrounding embedding medium (\cref{fig:thermodynamics}c).
The equilibrium condition for the strain yields,
\begin{equation}
    \left.\frac{\partial g}{\partial\tt{\eta}}\right|_{\tt{\eta}^\text{eq}}
    =
    0
    \quad\text{or}
    \quad
    \tt{\Pi}
    =
    \frac{1}{\mathring{V}} \left. \frac{\partial A}{\partial \tt{\eta}}\right|_{\tt{\eta}^\text{eq}}
    \equiv
    \tt{\sigma}^{\eta}(\mathring{\tt{h}};\tt{\eta}^\text{eq}).
    \label{eq:stress}
\end{equation}
which lets us identify $\tt{\Pi}$ as the conjugate to the Green--Lagrange strain $\tt{\eta}$, so that $\tt{\Pi}$ has the properties of a  second Piola--Kirchhoff stress.
It is important to emphasize that $\tt{\Pi}$ is a property of the embedding medium while $\tt{\sigma}^{\eta}$ is a property of the system, and that equilibrium requires $\tt{\Pi} = \tt{\sigma}^{\eta}$, see \cref{fig:thermodynamics}b.
\Cref{eq:gibbs-finite-strain,eq:stress} depend parametrically on the reference configuration $\mathring{\tt{h}}$, as the strain $\tt{\eta}$ is defined with respect to this reference (see \cref{fig:thermodynamics}a and b).
Of course, this equilibrium is constrained by the conditions described in \cref{sec:thermodynamic-potentials}, in particular that the cell remains a parallelepiped.

We can remove the choice of an arbitrary reference state in \cref{eq:stress} by choosing the h-matrix that corresponds to $\tt{\eta}^\text{eq}$ as our new reference $\mathring{\tt{h}}$.
For this choice, $\tt{\eta}^\text{eq}=0$.
The explicit dependence on $\tt{\eta}^\text{eq}$ in \cref{eq:stress} then disappears, and we can write
\begin{equation}
    \tt{\Pi}(\mathring{\tt{h}})
    =
    \frac{1}{\mathring{V}} \left. \frac{\partial A}{\partial \tt{\eta}}\right|_{\tt{\eta}=0}.
\end{equation}
This corresponds to how we typically estimate stresses in molecular calculations: We pick an h-matrix $\mathring{\tt{h}}$ and then perform small perturbations of the cell to determine the stress.

The derivative $\partial g/\partial \tt{\eta}$ must be a monotonously-increasing function passing through zero at $\tt{\eta}^\text{eq}\equiv 0$.
Loss of monotonous increase of this function means loss of convexity of the extended Gibbs free energy, and hence loss of (mechanical) stability.
We \emph{define} the elastic constants $\ff{\cc}$ as the second derivative of the extended Gibbs free energy
\begin{equation}
    \ff{\cc}^{\Pi}(\tt{\Pi};\mathring{h})
    \equiv
    \frac{1}{\mathring{V}}
    \left.
    \frac{\partial^2 g}{\partial \tt{\eta}^2}
    \right|_{\tt{\eta}=0},
    \label{eq:c_extended_gibbs}
\end{equation}
where the superscript on the elastic constant tensor $\ff{\cc}$ indicates the stress that was held constant.
Evaluating the elastic constants at constant $\tt{\Pi}$ yields
\begin{equation}
    \ff{\cc}^{\Pi}
    =
    \frac{1}{\mathring{V}}
    \left.
    \frac{\partial^2 A}{\partial \tt{\eta}^2}
    \right|_{\tt{\eta}=0}
    =
    \left.
    \frac{\partial \tt{\sigma}^{\eta}}{\partial \tt{\eta}}
    \right|_{\tt{\eta}=0}.
    \label{eq:elastic-constants-Seta}
\end{equation}
This expressions shows directly that the elastic constants are the second derivative of the free energy -- and expression that we might have written down intuitively and that is often the starting point in related works.
Note that we can similarly \emph{define}
\begin{equation}
    \ff{\cc}^{\prime}
    =
    \left.
    \frac{1}{\mathring{V}}
    \frac{\partial^2 A}{\partial \tt{\varepsilon}^2}
    \right|_{\tt{\varepsilon}=0},
    \label{eq:cepsilon}
\end{equation}
but this elastic tensor does not equal $\ff{\cc}^{\Pi}$.
\cref{eq:epsilon-to-eta} can be used to relate $\tt{\eta}$ and $\tt{\varepsilon}$
\begin{equation}
    \cc^{\prime}_{\alpha\beta\mu\nu}
    =
    \cc^{\Pi}_{\alpha\beta\mu\nu}
    +
    \sigma_{\alpha\mu} \delta_{\beta\nu}.
\end{equation}

Conceptually, we made small deformations of the equilibrium configuration and measure the resulting changes in stress (see \cref{fig:thermodynamics}c) at constant second Piola--Kirchhoff stress $\tt{\Pi}$.
We are often interested in an embedding medium that sustains a constant Cauchy rather than Piola--Kirchhoff stress.
We use \cref{eq:workincrcauchy} and the pull-back \cref{eq:pk2cauchy} to write the deviation of the extended Gibbs free energy from equilibrium as
\begin{equation}
    \delta g = \delta A - \mathring{V} J (\tt{F}^{-1} \cdot \tt{\sigma} \cdot \tt{F}^{-T}):\delta\tt{\eta}.
    \label{eq:dg-constant-Cauchy-stress}
\end{equation}
Unfortunately, this equation is not a differential, i.e. we cannot write a potential $g$ that yields \cref{eq:dg-constant-Cauchy-stress} as a derivative.
This is most easily seen by computing the elastic constants at constant Cauchy stress,
\begin{equation}
\begin{split}
    \cc^{\sigma}_{\alpha\beta\mu\nu}
    &=
    \frac{1}{\mathring{V}}
    \left.
    \frac{\partial}{\partial \eta_{\alpha\beta}}
    \frac{\delta g}{\delta \eta_{\mu\nu}}
    \right|_{\tt{\eta}=0} \\
    &=
    \cc^{\Pi}_{\alpha\beta\mu\nu}
    -
    \sigma_{\alpha\beta} \delta_{\mu\nu}
    +
    \sigma_{\beta\nu} \delta_{\alpha\mu}
    +
    \sigma_{\alpha\nu} \delta_{\beta\mu}.
    \label{eq:elnotsym}
\end{split}
\end{equation}
Because $\tt{\eta}$ is symmetric, $\eta_{\alpha\beta}\equiv \eta_{\beta\alpha}$ for $\alpha\not=\beta$, we can symmetrize \cref{eq:elnotsym} with respect to $\mu \leftrightarrow \nu$ (it is already symmetric in $\alpha \leftrightarrow \beta$), yielding
\begin{equation}
\begin{split}
    \cc^{\sigma}_{\alpha\beta\mu\nu}
    =
    \cc^{\Pi}_{\alpha\beta\mu\nu}
    +
    \frac{1}{2}
    (
    &\sigma_{\alpha\mu} \delta_{\beta\nu}
    +
    \sigma_{\alpha\nu} \delta_{\beta\mu}
    +
    \sigma_{\beta\mu} \delta_{\alpha\nu} \\
    &+
    \sigma_{\beta\nu} \delta_{\alpha\mu}
    -
    2\sigma_{\alpha\beta} \delta_{\mu\nu}
    ).
    \label{eq:elsym}
\end{split}
\end{equation}
If the elastic constants were given as the second derivative of a thermodynamic potential, then by virtue of the symmetry of second derivatives (Schwarz theorem) they must fulfill Voigt symmetry, namely
\begin{equation}
    \cc_{\alpha\beta\mu\nu} 
    = 
    \cc_{\beta\alpha\mu\nu}
    = 
    \cc_{\alpha\beta\nu\mu}
    = 
    \cc_{\mu\nu\alpha\beta}.
\end{equation}
Indeed, this symmetry is fulfilled by $\ff{\cc}^{\Pi}$ and $\ff{\cc}^{\prime}$, but not by $\ff{\cc}^{\sigma}$.

The elastic tensors at constant Cauchy stress $\ff{c}^{\sigma}$, at constant second Piola--Kirchhoff stress $\ff{c}^{\Pi}$ and $\ff{c}^\prime$ hence differ by offsets that explicitly depend on the Cauchy stress.
The elements of $\ff{c}^{\Pi}$ are sometimes called the \emph{Birch coefficients}~\citep{birch_finite_1947,birch_elasticity_1952,barron_second-order_1965,wallace_thermoelasticity_1967,wang_crystal_1993}.
The elements of $\ff{c}^\prime$ govern wave propagation~\citep{thurston_wave_1965,wallace_thermoelasticity_1967}.

\begin{figure}
    \begin{centering}
    \includegraphics[width=\columnwidth]{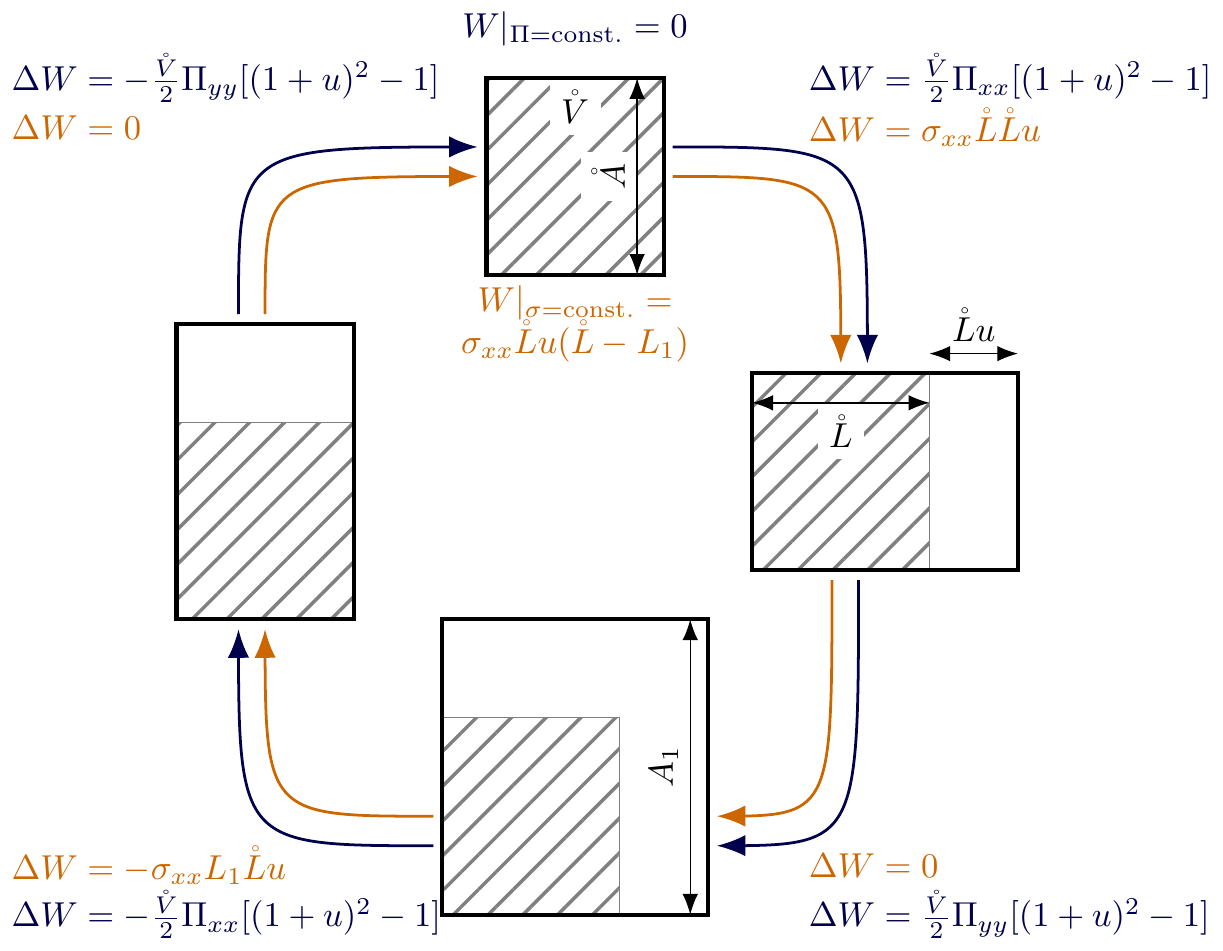}
	\caption{\label{fig:path_dependence}Mechanical work required to deform an RVE along a simple, exemplary closed deformation path while maintaining either the Cauchy stress $\tt{\sigma}$ or the Piola--Kirchhoff stress $\tt{\Pi}$ constant. At constant $\tt{\sigma}$ (black), the work around this cycle is not zero while it is zero at constant $\tt{\Pi}$ (orange). The reason for this is that when extending the area $\mathring{A}$ to $A_1$, the effective force on this area increases at constant $\tt{\sigma}$ but is constant at constant $\tt{\Pi}$. This path dependence implies that the state of the bath has changed after this cycle at constant $\tt{\sigma}$.} 
    \end{centering}
\end{figure}

The fact that we cannot write a thermodynamic potential for constant Cauchy stress means, that the work on the system depends on the deformation path.
This is illustrated in \cref{fig:path_dependence}, which compares the work at constant Cauchy stress with a constant second-Piola-Kirchhoff-stress.
As should become clear from this illustration, the reason for this path dependence is that in order to maintain a constant Cauchy stress, the embedding medium needs to adjust the stress on the system to its shape, e.g., by applying a constant force on the $yz$ plane and another constant force on the $xz$ plane.
In the context of molecular dynamics simulations, this means that extended system methods for simulations at constant non-isotropic Cauchy stress cannot be formulated~\citep{miller_molecular_2016}.
The classical Parinello--Rahman method~\citep{parrinello_polymorphic_1981} maintains constant $\tt{\Pi}$ and controlling the \emph{Cauchy} stress requires non-conservative methods~\citep{miller_molecular_2016}.

The Birch coefficients fulfill Voigt symmetry under purely hydrostatic pressure $P$ where $\sigma_{\alpha \beta} = -P \delta_{\alpha \beta}$~\citep{barron_second-order_1965}.
This path dependence hence disappears in the classical thermodynamic treatment that only considers changes in volume, as outlined in \cref{sec:thermodynamic-potentials}.

\subsection{Non-affine displacements}
The assumption that the deformation of an RVE is purely homogeneous and affine cannot be generally transferred to a (continuum) micromechanical or atomistic description of solids.
In most systems, atoms experience forces in the affinely deformed state that drive them into a new equilibrium position.
To capture this effect, we modify \cref{eq:affine} and write the new equilibrium positions as
\begin{equation}
    \vv{r}_i = \tt{F} \cdot \mathring{\vv{r}}_i + \vv{\chi}_i
\end{equation}
where $\vv{\chi}_i$ describes the displacement from the affine position of atom $i$ and is called the \emph{non-affine displacement}.
For convenience, we will work below with scaled coordinates $\vv{s}_i$, defined such that
\begin{equation}
    \vv{r}_i = \tt{h} \cdot (\mathring{\vv{s}}_i + \vv{\Psi}_i),
\end{equation}
with the scaled non-affine displacements $\vv{\Psi}_i$.
This allows us to decouple variation of (scaled) positions and variation of strain.

To understand the effect of non-affine displacements, we now turn to a variational formulation of the elastic constants, as for example employed in stochastic micromechanical homogenization~\citep{kozlov_averaging_1980,papanicolaou_boundary_1981} seeking effective elastic constants for an (elastically) heterogeneous medium.
The variational formulation requires a thermodynamic potential and we can only carry it out at constant second Piola--Kirchhoff stress $\tt{\Pi}$.
We search for the elastic constants $\ff{\CC}^{\Pi}$, such that
\begin{equation}
    \frac{1}{2}
    \mathring{V}
    \eta_{\alpha\beta}
    \CC^{\Pi}_{\alpha\beta\mu\nu}
    \eta_{\mu\nu}
    =
    \min_{\{\vv{\Psi}_i\}} g(\{\vv{\Psi}_i\}, \tt{\Pi}; \tt{\eta})
    \label{eq:zerostressnon-affine}
\end{equation}
for suitably small $\tt{\eta}$.
Without non-affine displacements ($\vv{\Psi}_i=0$) this immediately leads to the (affine) elastic constant $\ff{\cc}^{\Pi}$ given by \cref{eq:elsym}.

We expand the extended Gibbs free energy, i.e., the right-hand side of \cref{eq:zerostressnon-affine}, to second order in both strain and non-affine displacements around the equilibrium positions $\vv{\Psi}_i=0$.
Note that $\partial g/\partial \vv{\Psi}_i=0$ at equilibrium, otherwise the atoms would move without perturbation of the RVE.
This yields
\begin{equation}
\begin{split}
    g(\{\vv{\Psi}_i\}, \tt{\Pi}; \tt{\eta})
    =
    &\frac{1}{2}
    \mathring{V}
    \eta_{\alpha\beta}
    \cc^{\Pi}_{\alpha\beta\mu\nu}
    \eta_{\mu\nu} \\
    &+
    \frac{1}{2}
    \v{\Psi}
    \cdot
    \t{H}
    \cdot
    \v{\Psi}
    +
    \v{\Psi} \cdot
    \v{\Gamma}_{\alpha\beta}
    \eta_{\alpha\beta},
    \label{eq:gibbsexpanded}
\end{split}
\end{equation}
with Hessian $\t{H}=\partial^2 U/\partial\v{\Psi} \partial\v{\Psi}$ and non-affine forces $\v{\Gamma}_{\alpha\beta}=\partial^2 U/\partial\v{\Psi} \partial \eta_{\alpha\beta}$.
Boldface symbols (e.g. $\v{\Psi}$ or $\v{\Gamma}$) indicate $DN$-vectors that combine the individual $D$-vectors ($\vv{\Psi}_i$, $\vv{\Gamma}_i$) for all atoms and bold open symbols (e.g. $\t{H}$) indicate $DN\times DN$ matrices containing the $D\times D$ matrices ($\tt{H}_{ij}$) as blocks. Minimizing \cref{eq:gibbsexpanded} with respect to $\v{\Psi}$ gives
$\v{\Psi}
    =
    -
    \t{H}^{-1}
    \cdot
    \v{\Gamma}_{\alpha\beta}
    \eta_{\alpha\beta}$
and inserting this solution into \cref{eq:gibbsexpanded} yields
\begin{equation}
\begin{split}
    g(\{\vv{\Psi}_i\}, \tt{\Pi}; \tt{\eta})
    =
    &\frac{1}{2}
    \mathring{V}
    \eta_{\alpha\beta}
    \cc^{\Pi}_{\alpha\beta\mu\nu}
    \eta_{\mu\nu} \\
    &-
    \frac{1}{2}
    \eta_{\alpha\beta}
    \v{\Gamma}_{\alpha\beta}
    \cdot
    \t{H}^{-1}
    \cdot
    \v{\Gamma}_{\mu\nu}
    \eta_{\mu\nu}.
\end{split}
\end{equation}
Comparing this with the left hand side of \cref{eq:zerostressnon-affine}, we see that the finite stress elastic constants, including the effect of non-affine displacements, are given by
\begin{equation}
    \CC^{\Pi}_{\alpha\beta\mu\nu}
    =
    \cc^{\Pi}_{\alpha\beta\mu\nu}
    -
    \frac{1}{\mathring{V}}
    \v{\Gamma}_{\alpha\beta}
    \cdot
    \t{H}^{-1}
    \cdot
    \v{\Gamma}_{\mu\nu}.
    \label{eq:non-affineelastic}
\end{equation}
\Cref{eq:non-affineelastic} is identical to what has been reported in the literature for the unstressed case~\citep{lutsko_generalized_1989,lemaitre_sum_rules_2006,karmakar_athermal_2010}.
\Cref{eq:non-affineelastic} with $\ff{C}^{\Pi}$ replaced by $\ff{C}^{\sigma}$ also holds for the Birch coefficients since the Cauchy stress is not affected by non-affine displacements.

\subsection{\label{sec: lattice_stability}Stability criteria}

If a solid is subjected to mechanical load, it can become unstable and undergo either a polymorphic phase transition or -- for crystals -- transform into an amorphous state~\citep{machon_PIA_AAT_2014}.
The RVE remains mechanically stable as long as small perturbations of either atomic positions or h-matrix do not lead to a decrease in energy.
This is equivalent to stating that all eigenvalues of the total Hessian of the RVE, given by the matrix
\begin{equation}
    \mathcal{H}
    =
    \begin{pmatrix}
    \t{H} & \v{\tt{\Gamma}}^T \\
    \v{\tt{\Gamma}} & \ff{c}^\Pi
    \end{pmatrix}
    \label{eq:global-hessian}
\end{equation}
of second derivatives of the extended Gibbs free energy $g(\v{\Psi}, \tt{\eta})$, are nonnegative.
This is equivalent to stating, that the quadratic form
\begin{equation}
    \begin{pmatrix}
        \v{\Psi} \\ \tt{\eta}
    \end{pmatrix}
    \cdot
    \begin{pmatrix}
        \t{H} & \v{\tt{\Gamma}}^T \\
        \v{\tt{\Gamma}} & \ff{c}^\Pi
    \end{pmatrix}
    \cdot
    \begin{pmatrix}
        \v{\Psi} \\ \tt{\eta}
    \end{pmatrix}
    \label{eq:quadratic-form-displacement}
\end{equation}
must be nonnegative for small $\v{\Psi}$ and $\tt{\eta}$.
We can write a related quadratic form in terms of the forces $\v{f}$ and stresses $\tt{\sigma}$,
\begin{equation}
    \begin{pmatrix}
        \v{f} \\ \tt{\sigma}
    \end{pmatrix}
    \cdot
    \begin{pmatrix}
        \t{H} & \v{\tt{\Gamma}}^T \\
        \v{\tt{\Gamma}} & \ff{c}^\Pi
    \end{pmatrix}^{-1}
    \cdot
    \begin{pmatrix}
        \v{f} \\ \tt{\sigma}
    \end{pmatrix},
    \label{eq:quadratic-form-forces}
\end{equation}
which also must be nonnegative for small $\v{f}$ and $\tt{\sigma}$.
\Cref{eq:quadratic-form-forces} has to be interpreted as the pseudo-inverse, as the overall Hessian is not formally invertible because of translational invariance of the system.

The variety of stability criteria found in the literature are most easily described using \cref{eq:quadratic-form-displacement,eq:quadratic-form-forces}.
Assuming a fixed cell ($\tt{\eta}=0$), the system remains stable as long as all normal mode frequencies (given by the eigenvalues of $\t{H}$) are positive.
This is known as the \emph{dynamical stability}~\citep{born_stability1_1940,wallace_stability_1965,mouhat_necessary_2014}.
Under fixed positions ($\v{\Psi}=0$), the system remains stable as long as the eigenvalues of $\ff{c}^\Pi$ are positive.
This is known as the \emph{Born stability criterion}\citep{born_stability1_1940,born_dynamical_1954,wallace_stability_1965,hill_elasticity_and_stability_1975,wang_crystal_1993,wang_mechanical_1995,wang_unifying_stability_2012}.
If we assume that the forces on all atoms vanish ($\v{f}=0$), the eigenvalues of the bottom right block of $\mathcal{H}^{-1}$ must vanish.
It is straightforward to show that this block is given by \cref{eq:non-affineelastic}, the elastic constant tensor $\ff{C}^\Pi$ that includes the effect of nonaffine displacements.
Violation of Born stability or stability under consideration of nonaffine displacements is called an \emph{elastic instability}~\citep{hill_elasticity_and_stability_1975,wang_crystal_1993,wang_mechanical_1995,wang_unifying_stability_2012,mouhat_necessary_2014}. 

There are a few subtleties to consider.
First, dynamical or elastic stability are only necessary, not sufficient conditions for stability.
This is because stability is governed by $\mathcal{H}$, while regarding either dynamical or elastic stability considers only one of the diagonal blocks $\t{H}$ and $\ff{c}^\Pi$.
Those couple through $\tt{\v{\Gamma}}$, and this coupling will generally decrease the lowest individual eigenvalue.
Second, stability depends on the strain measure that is used to describe cell deformation.
For example, using the small strain tensor $\tt{\varepsilon}$ in lieu of the Lagrange strain $\tt{\eta}$ for constructing the global Hessian $\mathcal{H}$ will lead to the occurrence of the elastic constant tensor $\ff{c}^\prime$ (defined in \cref{eq:cepsilon}) as the bottom right block of \cref{eq:global-hessian}.
The stability of a molecular calculation run with a Parinello--Rahman barostat, for example, requires $\ff{C}^\Pi$ to be positive definite.
In general, stability therefore depends on the type of elastic embedding medium used for the RVE and the boundary conditions on the boundary of this medium.
For example, stability of a solid near a crack tip will be different from stability in a bending beam.

Third, stability at constant Cauchy stress needs to be discussed separately.
It does not emerge from a systematic expansion of $g$, and as a consequence $\ff{C}^\sigma$ is not symmetric for multiaxial stress.
Elastic stability implies that the quadratic form
\begin{equation}
    \eta_{\alpha \beta} \CC^{\sigma}_{\alpha\beta\mu\nu} \eta_{\mu \nu},
\end{equation}
remains non-negative. 
We can decompose the tensor $\CC^{\sigma}_{\alpha\beta\mu\nu}$ into its symmetric $\CC^{\text{sym}}_{\alpha\beta\mu\nu}$ and antisymmetric part $\CC^{\text{asym}}_{\alpha\beta\mu\nu}$, where $\CC^{\text{sym}}_{\alpha\beta\mu\nu} = (\CC^{\sigma}_{\alpha\beta\mu\nu} + \CC^{\sigma}_{\mu\nu\alpha\beta})/2$ and  $\CC^{\text{asym}}_{\alpha\beta\mu\nu} = (\CC^{\sigma}_{\alpha\beta\mu\nu} - \CC^{\sigma}_{\mu\nu\alpha\beta})/2$.
This yields
\begin{equation}
    \eta_{\alpha \beta}
    \CC^{\text{sym}}_{\alpha\beta\mu\nu}
    \eta_{\mu \nu}
    +
    \eta_{\alpha \beta}
    \CC^{\text{asym}}_{\alpha\beta\mu\nu}
    \eta_{\mu \nu}.
\end{equation}
It is easy to see that the second expression is always zero, since $\CC^{\text{asym}}_{\alpha \beta\mu \nu} = -\CC^{\text{asym}}_{\mu\nu\alpha \beta}$.
The stability condition then requires $\CC^{\text{sym}}_{\alpha\beta\mu\nu}$ to be positive definite.
For crystals, it is possible to derive closed form expressions for this condition.
A summary of necessary and sufficient conditions for common crystal structures is given in Ref.~\citep{mouhat_necessary_2014}. 

Fourth, for solids in which the non-affine contribution to the elastic constants is not negligible, the dynamical and the elastic stability criteria are not independent of each other.
To see this, we perform a diagonalization of the Hessian $\t{H} = \t{S} \t{D} \t{S}^{-1}$.
Here, $\t{S} = (\v{e}_1,\ldots,\v{e}_{3N})$ is a square matrix whose $p$-th column is the eigenvector  $\v{e}_p$ of $\t{H}$ and $\t{D}$ is a diagonal matrix whose elements are the eigenvalues $\lambda_p$.
The Hessian evaluated at a local minimum is positive semidefinite, in particular $\lambda_p > 0$ (for $p>D$) while the first $D$ eigenvalues are zero.
We can now express the inverse Hessian $\t{H}^{-1}$ in \cref{eq:non-affineelastic} through its eigenvectors and eigenvalues,
\begin{equation}
    -
    \frac{1}{\mathring{V}}
    \v{\Gamma}_{\alpha\beta}
    \cdot
    \t{H}^{-1}
    \cdot
    \v{\Gamma}_{\mu\nu}
    =
    -
    \frac{1}{\mathring{V}}
    \sum_p 
    \frac{\hat{\Gamma}_{p, \alpha \beta} \hat{\Gamma}_{p, \mu \nu}}{\lambda_p}
    \label{eq:eigendecomposition_na}
\end{equation}
where we have implicitly excluded the translational degrees of freedom from the sum.
The quantity $\hat{\Gamma}_{p, \alpha \beta} = \v{\Gamma}_{\alpha\beta} \cdot \v{e}_p$ is the projection of the non-affine forces on the eigenvectors of the Hessian.
If one eigenvalue $\lambda$ is small (but finite) the non-affine contribution of this mode will be a large negative value, which may lead to at least one very small elastic tensor element, unless the projected non-affine forces disappear.
In crystals, most of the $\hat{\Gamma}_{p, \alpha \beta}$ are zero because of symmetry.
Nonzero $\hat{\Gamma}_{p, \alpha \beta}$ only occur for crystals which have more than one nonequivalent Wyckoff position.
This means in crystals, elastic instabilities are largely decoupled from dynamical instabilities.
Conversely, in amorphous materials $\hat{\Gamma}_{p, \alpha \beta}$ are generally nonzero and the instabilities are closely coupled.

Nevertheless, this relation does not necessarily imply that a dynamically unstable solid is also elastically unstable.
Consider, for example, a solid filled with gas molecules which can move freely, such as a zeolite filled with air~\citep{bouessel_thermal_2014}.
In this case, the Hessian matrix $\t{H}$ is certainly not positive definite which implies that the system is not dynamically stable (because the air molecules can move), yet the solid remains elastically stable.
In the context of random networks, stability is explained by the concept of rigidity percolation, and rigid networks can contain floppy regions without loosing overall stability~\cite{Thorpe1983-ft,He1985-rn,Phillips1985-ls}.

\section{\label{chap: many-body_analytical}Many-body interatomic potentials}
\subsection{\label{sec: functional_forms} Specific functional forms}

In order to compute the elastic constants analytically, we need the second-order derivatives of the potential energy with respect to strain and atomic positions.
This yields the Born elastic constants $\ff{\cc}^{\Pi}$, the non-affine forces $\v{\Gamma}_{\alpha\beta}$ and the Hessian $\t{H}$ of the underlying potential.
We here outline these expressions for many-body interatomic potentials.
We consider the generic functional form~\citep{muser_interatomic_2022}
\begin{equation}
    U 
    =
    \frac{1}{2}
    \sum_{\substack{ij\\ i\neq j}}
    U_2(r^2_{ij}) + U_\text{m}(r^2_{ij}, \xi_{ij})
    \label{eq:genU}
\end{equation}
with
\begin{equation}
    \xi_{ij} 
    = 
    \sum_{\substack{k\\ k\neq i,j}} 
    \Xi(r^2_{ij}, r^2_{ik}, r^2_{jk}).
    \label{eq:genxi}
\end{equation}
Here $r_{ij}=|\vv{r}_{ij}|$ is the distance between atoms $i$ and $j$.
The three-body term, \cref{eq:genxi}, describes the triplet through the three side lengths of the triangle that it forms.
We express everything in terms of the squares of the distances, which simplifies derivatives with respect to the Green--Lagrange strain.
This trick has first been used by Born \citep{born_stability1_1940,born_dynamical_1954}. 

Our formulation trivially includes pair potentials $V(r_{ij})$,
\begin{equation}
    U_2(r^2_{ij}) = V(r_{ij})
    \quad\text{and}\quad
    U_\text{m} = 0.
\end{equation}
Furthermore, we can represent potentials of the Abell--Tersoff--Brenner~\citep{abell_empirical_1985,tersoff_new_1986,brenner_empirical_1990} type. For example, the multicomponent Tersoff potential~\citep{tersoff_modeling_1989} is given by
\begin{equation}
\begin{split}
    U_2(r_{ij}^2)
    &=
    f_\mathrm{C}(r_{ij})
    A_{ij} \exp(-\lambda_{ij} r_{ij}) 
    \\
    U_\text{m}(r_{ij}^2, \xi_{ij})
    &=
    -
    f_\mathrm{C}(r_{ij})
    b_{ij}(\xi_{ij})
    %\chi_{ij}(1+(\beta_i\xi_{ij})^{n_i})^{-1/2n_i}
    B_{ij} \exp(-\mu_{ij} r_{ij})
    \label{eq: bond_order_potential}
\end{split}
\end{equation}
and
\begin{equation}
    \Xi(r^2_{ij}, r^2_{ik}, r^2_{jk})
    = 
    f_\mathrm{C}(r_{ik})
    \left[1+\frac{c_i^2}{d_i^2} - \frac{c_i^2}{d_i^2 + (h_i - \cos\theta_{ijk})^2}\right],
\end{equation}
with
\begin{equation}
\begin{split}
    b_{ij}(\xi_{ij})
    &=
    \chi_{ij}(1+(\beta_i\xi_{ij})^{n_i})^{-1/2n_i} \\
    f_\mathrm{C}(r)
    &=
    \begin{cases}
    1, & r \leq r_{1} \\
    \frac{1}{2}\left(1 + \cos\left(\pi \frac{r - r_{1}}{r_{2} - r_{1}}\right)\right), & r_{1} < r < r_{2} \\
    0, & r_{2} \leq r
    \end{cases}
\end{split}
\end{equation}
where $\cos\theta_{ijk}=(r^2_{ij}+r^2_{ik}-r_{jk}^2)/(2 r_{ij}r_{ik})$ and $f_\mathrm{C}$ is a cutoff function that varies smoothly from unity to zero between two distances.
The quantities $A_{ij}$, $B_{ij}$, $\lambda_{ij}$, $\mu_{ij}$, $\chi_{ij}$, $\beta_i$, $n_i$, $c_i$, $d_i$, $h_i$, $r_{1}$ and $r_{2}$ are parameters (see Ref.~\citep{tersoff_modeling_1989} for more information).

This generic functional form is also suitable for cluster potentials which include interactions only up to three-body terms.
One frequently used example of this class of potentials is the Stillinger--Weber potential, whose expressions in our formulation read~\citep{stillinger_computer_1985}
\begin{align}
    U_2(r_{ij}^2)
    &=
    f_{\mathrm{C}1}(r_{ij})
    A \epsilon
    \left[
    B \left(\frac{\sigma}{r_{ij}} \right)^p - \left( \frac{\sigma}{r_{ij}}\right)^q 
    \right] 
    \\
    U_\text{m}(r_{ij}^2, \xi_{ij})
    &=
    \epsilon \lambda \xi_{ij}
\end{align}
and
\begin{equation}
    \Xi(r^2_{ij}, r^2_{ik}, r^2_{jk})
    =  
    f_{\text{C}2}(r_{ik})
    f_{\text{C}2}(r_{ij})
    \left[
    \cos(\theta_{ijk}) - \cos(\theta_0) 
    \right]^2 
\end{equation}
with
\begin{equation}
\begin{split}
    f_{\text{C}1}(r)
    &=
    H(a\sigma - r)
    \exp\left(\frac{\sigma}{r - a\sigma}\right)
    \\
    f_{\text{C}2}(r)
    &=
    H(a\sigma - r)
    \exp\left(\frac{\gamma \sigma}{r - a\sigma}\right)
\end{split}
\end{equation}
where $H$ is the Heaviside step function, $\cos(\theta_0)=-1/3$ is the tetrahedral angle, while
$\sigma$, $\epsilon$, $A$, $B$, $p$, $q$, $a$, $\lambda$, and $\gamma$ are parameters.
The values for the original parametrization can be found in Ref.~\citep{stillinger_computer_1985}.

\subsection{Generic form of the first derivatives}

In the following we use the shorthand notation $R_{ij}=r^2_{ij}$ so that the square is not mistaken as a second-order derivative.
Three derivatives will appear repeatedly in the following. 
First, we need the derivative of $R_{X}$ with respect to the atomic positions
\begin{equation}
    \frac{\partial R_{X}}{\partial r_{n,\beta}}
    = 
    2 \pi_{X|n} r_{X,\beta} 
    \quad
    \text{and}
    \quad
    \frac{\partial^2 R_{X}}{\partial r_{m,\alpha} \partial r_{n,\beta}}   
    =
    2 \pi_{X|n} \pi_{X|m} \delta_{\alpha \beta} 
\end{equation}
with $X \in \{ij, ik, jk \}$ and $\pi_{ij|n}=\delta_{in}-\delta_{jn}$.
Second, we need the derivative of $R_{X}$ with respect to the Green--Lagrange strain tensor $\tt{\eta}$.
Using $R_{X}=\mathring{r}_{X, \alpha} (\delta_{\alpha \beta}+2\eta_{\alpha \beta}) \mathring{r}_{X,\beta}$, we obtain
\begin{equation}
    \frac{\partial R_{X}}{\partial \eta_{\alpha\beta}}
    =
    2 \mathring{r}_{X,\alpha} \mathring{r}_{X, \beta}
    \quad
    \text{and}
    \quad
    \frac{\partial^2 R_{X}}{\partial \eta_{\alpha \beta} \partial \eta_{\mu \nu}} = 0.
\end{equation}
Finally, we consider the mixed derivative with respect to the atomic positions and the Green--Lagrange strain which reads
\begin{equation}
    \frac{\partial^2 R_{X}}{\partial r_{n,\gamma}\partial \eta_{\alpha\beta}}
    =
    2 
    \pi_{X|n}
    \left(
    \mathring{r}_{X,\beta} \delta_{\gamma \alpha}
    +
    \mathring{r}_{X,\alpha} \delta_{\gamma \beta}    
    \right)
\end{equation}
Using these derivatives we can directly write down the expression for the forces
\begin{equation}
\begin{split}
    \vv{f}_{n}
    \equiv 
    - \frac{1}{2}
    \frac{\partial U}{\partial \vv{r}_{n}}
    =
    &-
    \sum_{\substack{ij\\ i\neq j}}
    \Bigg(
        \pi_{ij|n} 
        \left[
        \frac{\partial U_2}{\partial R_{ij}} + \frac{\partial U_\text{m}}{\partial R_{ij}}
        \right]
        \vv{r}_{ij} 
    \\
    &+
        \sum_{\substack{k\\ k\neq i,j}}
        \sum_X
        \pi_{X|n}
        \frac{\partial U_\text{m}}{\partial \xi_{ij}}
        \frac{\partial\Xi}{\partial R_X}
        \vv{r}_{X}
    \Bigg)
\end{split}
\end{equation}
and the stress tensor
\begin{equation}
\begin{split}
    \sigma^{U\eta}_{\alpha \beta}
    \equiv 
    \frac{1}{\mathring{V}}
    \frac{\partial U}{\partial \eta_{\alpha \beta}}
    =
    &\frac{1}{\mathring{V}}
    \sum_{\substack{ij\\ i\neq j}}
    \Bigg(
    \left[
    \frac{\partial U_2}{\partial R_{ij}} + \frac{\partial U_\text{m}}{\partial R_{ij}}
    \right]
    r_{ij,\alpha} r_{ij, \beta}
    \\
    &+
    \sum_{\substack{k\\ k\neq i,j}}
    \sum_X
    \frac{\partial U_\text{m}}{\partial \xi_{ij}}
    \frac{\partial\Xi}{\partial R_X}
    r_{X,\alpha} r_{X,\beta}
    \Bigg).
\end{split}
\end{equation}

\onecolumngrid
\subsection{Generic form of the second derivatives}

We need general second-order derivatives of the form
$\partial^2 U/\partial a\partial b$
where $a$ and $b$ can be: Components of the Green--Lagrange strain $\tt{\eta}$ (yielding the Born elastic constants), components of the position vector $\vv{r}$ (yielding the Hessian) or combinations of both (yielding the non-affine forces).
We now write these down for the generic form of the many-body potential given by \cref{eq:genU,eq:genxi}.
The potential-specific derivatives of the functions $U_2$, $U_\text{m}$ and $\Xi$ appearing in the various expressions are left as exercises to the reader; they can also be found in our implementation of the methods included in the software package \textsc{matscipy}~\citep{matscipy}.
Taking the second derivative of \cref{eq:genU} yields
\begin{equation}
\begin{split}
    \frac{\partial^2 U}{\partial a\partial b}
    =
    \frac{1}{2}
    \sum_{\substack{ij\\ i\neq j}}
    \Bigg\{
        &\underbrace{
        \left[ 
        \frac{\partial U_2}{\partial R_{ij}} + \frac{\partial U_\text{m}}{\partial R_{ij}}
        \right]
        \frac{\partial^2 R_{ij}}{\partial a \partial b}}_\text{term 1}
        +
        \underbrace{
        \left[ 
        \frac{\partial^2 U_2}{\partial R_{ij}^2} + \frac{\partial^2 U_\text{m}}{\partial R_{ij}^2}
        \right]
        \frac{\partial R_{ij}}{\partial a}
        \frac{\partial R_{ij}}{\partial b}}_\text{term 2}
        +
        \underbrace{
        \frac{\partial U_\text{m}}{\partial \xi_{ij}}
        \frac{\partial^2 \xi_{ij}}{\partial a \partial b}}_\text{term 3}\\
        &+
        \underbrace{ 
        \frac{\partial^2 U_\text{m}}{\partial R_{ij} \partial \xi_{ij}} 
        \left(\frac{\partial R_{ij}}{\partial a} \frac{\partial \xi_{ij}}{\partial b}
        +
        \frac{\partial \xi_{ij}}{\partial a} \frac{\partial R_{ij}}{\partial b}\right)}_\text{term 4}
        +
        \underbrace{
        \frac{\partial^2 U_\text{m}}{\partial \xi_{ij}^2} \frac{\partial \xi_{ij}}{\partial a} \frac{\partial \xi_{ij}}{\partial b}}_\text{term 5}
    \Bigg\}.
    \label{eq:gensecond}
\end{split}
\end{equation}
Furthermore, the first derivative of $\xi_{ij}$ can be expressed as
\begin{equation}
    \frac{\partial \xi_{ij}}{\partial a} 
    =
    \sum_{\substack{k\\ k\neq i,j}}
    \sum_X
    \frac{\partial \Xi(R_{ij},R_{ik},R_{jk})}{\partial R_X} \frac{\partial R_X}{\partial a} 
\end{equation}
and its second derivative is given by 
\begin{equation}
\begin{split}
    \frac{\partial^2 \xi_{ij}}{\partial a \partial b} 
    =
    &\sum_{\substack{k\\ k\neq i,j}}
    \sum_{XY}
    \frac{\partial R_X}{\partial a}
    \frac{\partial^2 \Xi(R_{ij},R_{ik},R_{jk})}{\partial R_X \partial R_Y} \frac{\partial R_Y}{\partial b}
    +
    \sum_{\substack{k\\ k\neq i,j}}
    \sum_{X}
    \frac{\partial \Xi(R_{ij},R_{ik},R_{jk})}{\partial R_X}
    \frac{\partial^2 R_X}{\partial b \partial a}
\end{split}
\end{equation}
with $X,Y\in\{ij,ik,jk\}$.
In the following subsections, we provide analytic expressions for the Born elastic constants, the non-affine forces, and the Hessian.
Each of these expressions along with the interatomic potentials from \cref{sec: functional_forms} were tested for correctness against finite differences. 

\subsection{Born elastic constants}

For the calculation of the (Born) elastic constants, we need to evaluate $\mathring{V}\cc^{\Pi\eta}_{\alpha\beta\mu\nu}=\partial^2 U/\partial \eta_{\alpha\beta}\partial \eta_{\mu\nu}$,
hence both $a$ and $b$ are components of the Green--Lagrange strain tensor $\tt{\eta}$.
It is then straightforward to evaluate the individual terms for the Born elastic constants,
\begin{align}
    \mathring{V} \cc_{\alpha\beta\mu\nu}^{\Pi\eta(1)}
    &=
    0,
    \\
    \mathring{V} \cc_{\alpha\beta\mu\nu}^{\Pi\eta(2)}
    &=
    2
    \sum_{\substack{ij\\ i\neq j}}
    \left[
    \frac{\partial^2 U_2}{\partial R_{ij}^2}
    +
    \frac{\partial^2 U_\text{m}}{\partial R_{ij}^2}
    \right]
    \mathring{r}_{ij,\alpha} \mathring{r}_{ij,\beta} \mathring{r}_{ij,\mu} \mathring{r}_{ij,\nu},
    \\
    \mathring{V} \cc_{\alpha\beta\mu\nu}^{\Pi\eta(3)}
    &=
    2 
    \sum_{\substack{ij\\ i\neq j}}
    \sum_{\substack{k\\ k\neq i,j}}
    \sum_{XY} \frac{\partial U_\text{m}}{\partial \xi_{ij}} \frac{\partial^2\Xi}{\partial R_X \partial R_Y} 
    \mathring{r}_{X,\alpha} \mathring{r}_{X,\beta} \mathring{r}_{Y,\mu} \mathring{r}_{Y,\nu},
    \\
    \mathring{V} \cc_{\alpha\beta\mu\nu}^{\Pi\eta(4)}
    &=
    2 
    \sum_{\substack{ij\\ i\neq j}}
    \sum_{\substack{k\\ k\neq i,j}}
    \sum_{X} \frac{\partial^2 U_\text{m}}{\partial R_{ij} \partial \xi_{ij}}
    \frac{\partial \Xi}{\partial R_X}
    \left(
    \mathring{r}_{ij,\alpha} \mathring{r}_{ij,\beta} \mathring{r}_{X,\mu} \mathring{r}_{X,\nu} 
    +
    \mathring{r}_{X,\alpha} \mathring{r}_{X,\beta} \mathring{r}_{ij,\mu} \mathring{r}_{ij,\nu}
    \right),
    \\
    \mathring{V} \cc_{\alpha\beta\mu\nu}^{\Pi\eta(5)}
    &=
    2 
    \sum_{\substack{ij\\ i\neq j}}
    \frac{\partial^2 U_\text{m}}{\partial \xi_{ij}^2}
    \left(
        \sum_{\substack{k\\ k\neq i,j}}
        \sum_{X} 
        \frac{\partial \Xi}{\partial R_X} 
        \mathring{r}_{X,\alpha} \mathring{r}_{X,\beta}
    \right)
    \left(
        \sum_{\substack{l\\ l\neq i,j}}
        \sum_{Y} 
        \frac{\partial \Xi}{\partial R_Y}
        \mathring{r}_{Y,\mu} \mathring{r}_{Y,\nu}
    \right)
\end{align}
with $X\in\{ij,ik,jk\}$ and with $Y\in\{ij,il,jl\}$.

\subsection{Non-affine forces}
For the non-affine forces, $b$ becomes a component of an atomic position $r_{n,\gamma}$ and $a$ needs to be replaced with a component of the strain tensor $\eta_{\alpha\beta}$.
The five individual terms of \cref{eq:gensecond} yield
\begin{align}
    \Gamma^{(1)}_{n\gamma,\alpha\beta}
    =&
    \sum_{\substack{ij\\ i\neq j}}
    \pi_{ij|n}
    \left[ 
    \frac{\partial U_2}{\partial R_{ij}} + \frac{\partial U_\text{m}}{\partial R_{ij}}
    \right]
    \left(
    \mathring{r}_{ij,\beta} \delta_{\alpha \gamma} + \mathring{r}_{ij,\alpha} \delta_{\beta\gamma}
    \right),
    \\
    \Gamma^{(2)}_{n\gamma,\alpha\beta}
    =&
    2
    \sum_{\substack{ij\\ i\neq j}}
    \pi_{ij|n}
    \left[ 
    \frac{\partial^2 U_2}{\partial R_{ij}^2} + \frac{\partial^2 U_\text{m}}{\partial R_{ij}^2}
    \right]
    \mathring{r}_{ij,\alpha} \mathring{r}_{ij,\beta}
    \mathring{r}_{ij, \gamma},
    \\
    \Gamma^{(3)}_{n\gamma,\alpha\beta}
    =&
    2
    \sum_{\substack{ij\\ i\neq j}}
    \sum_{\substack{k\\ k\neq ij}}
    \sum_{XY}
    \pi_{Y|n}
    \frac{\partial U_\text{m}}{\partial \xi_{ij}}
    \frac{\partial^2 \Xi}{\partial R_X\partial R_Y}
    \mathring{r}_{X,\alpha} \mathring{r}_{X,\beta}
    \mathring{r}_{Y, \gamma}
    \notag\\
    &+
    \sum_{\substack{ij\\ i\neq j}}
    \sum_{\substack{k\\ k\neq ij}}
    \sum_{X}
    \frac{\partial U_\text{m}}{\partial \xi_{ij}}
    \frac{\partial \Xi}{\partial R_X}
    \pi_{X|n}
    \left(
    \mathring{r}_{X,\beta} \delta_{\alpha \gamma}
    +
    \mathring{r}_{X,\alpha} \delta_{\beta \gamma}
    \right),
    \\
    \Gamma^{(4)}_{n\gamma,\alpha\beta}
    =&
    2
    \sum_{\substack{ij\\ i\neq j}}
    \sum_{\substack{k\\ k\neq ij}}
    \sum_{X}
    \frac{\partial^2 U_\text{m}}{\partial R_{ij} \partial \xi_{ij}}
    \frac{\partial \Xi}{\partial R_X}
    \left(
        \pi_{ij|n}
        \mathring{r}_{X,\alpha}\mathring{r}_{X,\beta}
        \mathring{r}_{ij, \gamma}
        +
        \pi_{X|n}
        \mathring{r}_{ij,\alpha}\mathring{r}_{ij,\beta}
        \mathring{r}_{X, \gamma}
    \right),
    \\
    \Gamma^{(5)}_{n\gamma,\alpha\beta}
    =&
    2
   \sum_{\substack{ij\\ i\neq j}}
    \frac{\partial^2 U_\text{m}}{\partial \xi_{ij}^2}
   \left(
   \sum_{\substack{k\\ k\neq ij}}
   \sum_X
   \frac{\partial\Xi}{\partial R_X}
   \mathring{r}_{X,\alpha}\mathring{r}_{X,\beta}
   \right)
   \left(
   \sum_{\substack{l\\ l\neq ij}}
   \sum_Y
   \pi_{Y|n}
   \frac{\partial\Xi}{\partial R_Y}
   \mathring{r}_{Y, \gamma}   
   \right)
\end{align}
with $X\in\{ij,ik,jk\}$ and $Y\in\{ij,il,jl\}$.

\subsection{Hessian}

Starting again from \cref{eq:gensecond} and replacing $a$ as well as $b$ with components of actual atomic positions $r_{m,\alpha}$ and $r_{n,\beta}$ yields the individual expressions for the off-diagonal components of the Hessian.
Note that the expressions given here differ from the components of $\t{H}$, which are the derivatives with respect to scaled positions.
The (second) derivatives with respect to actual positions are given by
\begin{align}
    H^{(1)}_{m\alpha,n\beta}
    =&
   \sum_{\substack{ij\\ i\neq j}}
    \tau_{ij,ij|mn}
    \left[ 
    \frac{\partial U_2}{\partial R_{ij}} + \frac{\partial U_\text{m}}{\partial R_{ij}}
    \right]
    \delta_{\alpha \beta},
    \\
    H^{(2)}_{m\alpha,n\beta}
    =&
    2
    \sum_{\substack{ij\\ i\neq j}}
    \tau_{ij,ij|mn}
    \left[ 
    \frac{\partial^2 U_2}{\partial R_{ij}^2} + \frac{\partial^2 U_\text{m}}{\partial R_{ij}^2}
    \right]
    r_{ij,\alpha}
    r_{ij,\beta},
    \\
    H^{(3)}_{m\alpha, n\beta}
    =&
    2
   \sum_{\substack{ij\\ i\neq j}}
   \sum_{\substack{k\\ k\neq ij}}
    \sum_{XY}
    \tau_{X,Y|mn}
    \frac{\partial U_\text{m}}{\partial \xi_{ij}}
    \frac{\partial^2 \Xi}{\partial R_X \partial R_Y}
    r_{X,\alpha} r_{Y,\beta}\\
    &+
   \sum_{\substack{ij\\ i\neq j}}
   \sum_{\substack{k\\ k\neq ij}}
    \sum_{X}
    \tau_{X,X|mn}
    \frac{\partial U_\text{m}}{\partial \xi_{ij}}
    \frac{\partial \Xi}{\partial R_X}
    \delta_{\alpha \beta},
    \\
    H^{(4)}_{m\alpha, n\beta}
    =&
    2
   \sum_{\substack{ij\\ i\neq j}}
   \sum_{\substack{k\\ k\neq ij}}
    \sum_X
    \frac{\partial^2 U_\text{m}}{\partial R_{ij} \partial \xi_{ij}}
    \frac{\partial \Xi}{\partial R_{X}}
    \left(
        \tau_{X,ij|mn}
        r_{X,\alpha} r_{ij, \beta}
        +
        \tau_{ij,X|mn}
        r_{ij, \alpha} r_{X, \beta}
    \right),
    \\
    H^{(5)}_{m\alpha,n\beta}
    =&
    2
   \sum_{\substack{ij\\ i\neq j}}
    \frac{\partial^2 U_\text{m}}{\partial \xi_{ij}^2}
    \left(
    \sum_{\substack{k\\ k\neq ij}}
    \sum_X
     \pi_{X|m}
     \frac{\partial \Xi}{\partial R_X}
     r_{X,\alpha}
    \right)
    \left(
   \sum_{\substack{l\\ l\neq ij}}
    \sum_Y
     \pi_{Y|n}
     \frac{\partial \Xi}{\partial R_Y}
     r_{Y,\beta}
    \right),
\end{align}
with $X\in\{ij,ik,jk\}$ and $Y\in\{ij,il,jl\}$.
Furthermore, we have introduced the shorthand notation $\tau_{X,Y|mn}=\pi_{X|m}\pi_{Y|n}$.
We know that, by virtue of the symmetry of second derivatives, the full Hessian must be symmetric
\begin{equation}    
    H_{m\alpha,n\beta} = H_{n\beta,m\alpha}.
\end{equation}
We can relate the diagonal and off-diagonal elements of the Hessian with translational invariance. For any translation vector $d_\alpha$, we have, defining the uniform vector $a_m\equiv 1$,
\begin{align}
  0 & = \frac{\partial}{\partial d_\gamma} U(\{a_m d_\alpha\})\\
  & = \frac{\partial}{\partial d_\gamma}\left[ U({0}) + f_{m,\alpha}a_m d_\alpha
    + \frac{1}{2}a_m d_\alpha H_{m\alpha,n\beta} a_n d_\beta\right]\\
    & = f_{m,\gamma}a_m + a_mH_{m\gamma,n\beta}a_nd_\beta\\
  & = a_mH_{m\gamma,n\beta}a_nd_\beta,
\end{align}
where we have assumed that conservation of linear momentum $\sum_{m}f_{m,\gamma}=0$ holds.
The remaining condition is that $a_m$ is an eigenvector associated with a null eigenvalue of the Hessian, which translates to
\begin{align}
  H_{m\alpha,m\beta}
  =
  - \sum_{n\neq m}H_{m\alpha,n\beta}.
  \label{eq:diagonal}
\end{align}

\twocolumngrid
\section{\label{sec: methods} Methods}
To compute elastic properties and analyze lattice stability, it is necessary to first define materials, their specific potential parameterization, initial conditions, employed boundary conditions, and further computational methodology. 
We have chosen diamond cubic silicon, 3C silicon carbide and the $\alpha$-quartz form of silicon dioxide because they are ubiquitous in nature and have tremendous importance for technical applications in electronics, photonics, and photovoltaics \citep{jalali_silicon_photonics_2006,kimoto_siliconcarbide_2014,she_review_SiC_2017,heaney_silica_2018}. 
In the case of silicon, we consider bond-order potentials as well as one cluster potential~\citep{muser_interatomic_2022}.
We denote Tersoff's potential~\citep{tersoff_modeling_1989} as \textsc{TIII}, the refitted version for pure silicon of \citet{erhart_analytical_2005} as \textsc{EAII}, the cluster potential by \citet{stillinger_computer_1985} as \textsc{SW} and the modified Tersoff potential from \citet{kumagai_development_2007} as \textsc{Kum}.
The \textsc{Kum} potential differs from the TIII potential by the angular-dependent term and a modified cutoff function. 
We use the parametrization referred to as \textsc{MOD} in the publication of \citet{kumagai_development_2007}.
Similarly for silicon carbide, we denote the original parametrization of \citet{tersoff_modeling_1989} as TIII and its refitted version by \citet{erhart_analytical_2005} as EA.  
For silicon dioxide we employ the potential by \citet{vanBeest_BKS_1990}, which we refer to as \textsc{BKS}.
The short-range interaction is truncated and shifted at a cutoff distance of $r_c=10\,\text{\AA}$. 
Electrostatic interactions are treated using traditional Ewald summation \citep{ewald_original_1921,toukmaji_ewald_1996}.

The initial crystalline configurations are set up using the equilibrium lattice constant at zero temperature and pressure for the specific interatomic potential. 
The simulation cells contain $N\approx1000$ atoms with periodic boundary conditions in all spatial directions.
We first consider the case of hydrostatic pressure $\sigma^{\eta}_{\alpha\beta} = -P \delta_{\alpha\beta}$ in the limit of zero temperature.
We follow the convention that a compressing force creates a positive pressure $P > 0$ and causes a negative stress $\sigma < 0$.
Hydrostatic loading of the RVE is imposed by first applying an affine transformation $\tt{F} = (1 + \varepsilon)\tt{1}$ with a prescribed strain increment $\varepsilon = 10^{-3}$ to the simulation cell and the atoms and afterwards performing an energy minimization.

We note that large hydrostatic compression or tension can lead to nonphysical behavior for interatomic potentials with a finite interaction range.
The finite interaction range is typically implemented through a cutoff function making the potential energy and possibly its derivatives vanish at a distance $r_c$. 
The reason for this finite interaction is that many potentials are constructed on the assumption of only nearest-neighbor interactions, while further interactions are screened~\citep{Baskes1994-sq,Pastewka2008-he,pastewka_screened_2013,muser_interatomic_2022,muser_improved_cutoff_2022}.
The value of $r_c$ is then typically chosen to lie between first and second neighbor shell in the ground-state crystal.
Volumetric deformation, however, does not change nearest neighbor relationships, yet using a fixed finite interaction range will lead to unphysical zero-energy configurations at large volumetric strain.
For crystalline structures that do not change their bonding topology, a simple remedy is to retain nearest neighbor interactions up to arbitrary distances, effectively replacing the cutoff-procedure by a fixed bond topology~\citep{mizushima_ideal_1994,tang_thermomechanical_SiC_1995}.
In the following, we denote potentials where the bond topology is determined for the ideal crystal and kept fixed during deformation by appending \textsc{+FT} to the potential names defined in the last paragraph, e.g., for Tersoff's potential \textsc{TIII+FT}.

\begin{figure}[ht!]
    \centering
	\includegraphics[]{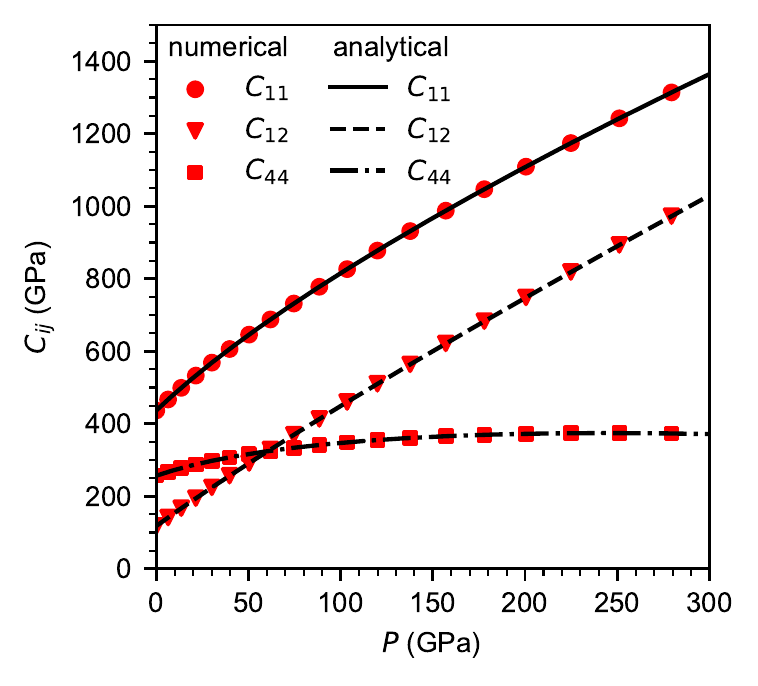}
	\caption{\label{fig: analytical_numerical} Numerical and analytical calculations of the elastic constants of silicon carbide as a function of pressure. Results are shown for the \textsc{TIII} interatomic potential with a fixed bonding topology.}
\end{figure}

For all employed interatomic potentials, we checked consistency of the analytical expressions from \cref{chap: many-body_analytical} with numerical results obtained from finite difference computations.
In \cref{fig: analytical_numerical} we show examplarily a comparison between numerical and analytical results for one interatomic potential for silicon carbide. 
Due to the small strain increment and the resulting large number of elastic constants, we replace the discrete values of the analytic results by a line.
From this figure it is evident that if proper parameters are chosen for the finite difference computation, both approaches to compute elastic constants lead to similar results.

\section{Results}
\label{sec:results}

\subsection{Elastic constants at zero stress}
We first compute the elastic constants at zero hydrostatic pressure.
This serves as a validation for the analytical expressions and enables us to determine, which interatomic potential best describes the elastic properties of the material under investigation. 
The different elastic constants are only equivalent in the limiting case of zero external stress~\citep{barron_second-order_1965}. 
We restrict our discussion to the Birch coefficients and will omit in the following the superscripts of the elastic constants even at nonzero external stress, $\CC_{\alpha\beta\mu\nu} \equiv \CC^{\sigma}_{\alpha\beta\mu\nu}$.
\Cref{Si_SiC_Quartz} shows a comparison of the equilibrium lattice constants and the elastic constants.
We compare experimental results, density functional theory (DFT) calculations, results from reference publications and our values from analytical calculations.
When reporting values, we condense the four indices of the elastic constant tensors to two using Voigt (or Nye) notation \cite{voigt1910,nye1985}.
Silicon and silicon carbide have a cubic lattice symmetry with three independent elastic constants $\CC_{11}$, $\CC_{12}$ and $\CC_{44}$. 
In contrast, $\alpha$-quartz has a trigonal crystal structure with six independent elastic moduli $\CC_{11}$, $\CC_{12}$, $\CC_{13}$, $\CC_{14}$, $\CC_{33}$ and $\CC_{44}$.

\begin{table*}[ht!]
\caption{\label{Si_SiC_Quartz} Zero-stress lattice constants and elastic constants for silicon, silicon carbide and silicon dioxide. $a_0$, $c_0$: Lattice constants; $\CC_{ij}$: elastic constants with non-affine contribution; $\cc_{ij}$: elastic constants without non-affine contribution.}
\begin{tabular}{l@{\extracolsep{2mm}}c@{\extracolsep{2mm}}c@{\extracolsep{1mm}}c@{\extracolsep{2mm}}c@{\extracolsep{1mm}}cc@{\extracolsep{2mm}}c}
%\begin{tabular}{lcccccc}
\multicolumn{8}{c}{Silicon}
\\\hline\hline
 & Expt. & \multicolumn{2}{c}{DFT-LDA} & \multicolumn{2}{c}{TIII} & \multicolumn{2}{c}{EA Si-II} 
\\
 & Ref.$^b$                            & Ref.$^k$     & Ref.$^c$& Ref.$^e$   & this work & Ref.$^a$     & this work\\
\hline
\\[-2mm]
$a_0$ (\AA)       & $5.431$ & $5.400$          & $5.406$ & $5.432$    & $5.432$  & $5.429$ & $5.432$ \\
$\CC_{11}$ (GPa)    & $166$   & $159$   & $160$   & $142.5$    & $142.54$ & $167$ & $168.48$ \\
$\CC_{12}$ (GPa)    & $64$    & $61$    & $63$    & $75.4$     & $75.38$  & $65$  & $63.47$ \\
$\CC_{44}$ (GPa)    & $80$    & $85$    & $82$    & $69$       & $69.03$  & $72$ &  $71.53$ \\
$\cc_{44}$ (GPa) &           & $111$   & $112$   & $118.8$    & $118.81$ & $111$ & $110.44$ 
\vspace{2mm}
\\
\multicolumn{5}{c}{}
\\\hline\hline
                  & \multicolumn{2}{c}{\textsc{Kum}} & \multicolumn{2}{c}{SW}
\\
                  & Ref.$^f$ & this work & Ref.$^l$ & this work \\
\hline
\hline
\\[-2mm]
$a_0$ (\AA)       & $5.429$ & $5.429$  & $5.431\phantom{^e}$  & $5.431$ \\
$\CC_{11}$ (GPa)    & $166.4$ & $166.37$ & $151.4\phantom{^e}$ & $151.42$ \\
$\CC_{12}$ (GPa)    & $65.3$  & $65.30$  & $76.4\phantom{^e}$ &  $76.42$ \\
$\CC_{44}$ (GPa)    & $77.1$  & $77.11$  & $56.4\phantom{^e}$ & $56.45$ \\
$\cc_{44}$ (GPa) & $120.9$ & $120.93$ & $117.2^e$ & $109.76$  
\vspace{5mm}
\\
\multicolumn{8}{c}{Silicon carbide}
\\\hline\hline
& Expt. & \multicolumn{2}{c}{DFT-LDA} & \multicolumn{2}{c}{TIII} & \multicolumn{2}{c}{EA}
\\
                   & Ref.$^m$& Ref.$^d$ & Ref.$^c$ & Ref.$^n$ & this work & Ref.$^a$ & this work\\
\hline
\\[-2mm]
$a_0$ (\AA)        & $4.3596$  & $4.344$      & $4.338$  & $4.32\phantom{^m}$  & $4.321$  & $4.359$ & $4.359$ \\
$\CC_{11}$ (GPa)   & $390$     & $390$   & $405$    & $420.0\phantom{^m}$ & $436.63$ & $382$   & $383.78$ \\
$\CC_{12}$ (GPa)   & $142$     & $134$   & $145$    & $120.0\phantom{^m}$ & $117.94$ & $145$   & $144.41$ \\
$\CC_{44}$ (GPa)   & $256$     & $253$   & $247$    & $260.0\phantom{^m}$ & $256.60$ & $240$   & $239.75$\\
$\cc_{44}$ (GPa)   & $-$       & $273$   & $311.0$    &  $311.0$    & $310.81$  & $305$   & $304.75$ 
\vspace{5mm}
\\
\multicolumn{8}{c}{Silicon dioxide} 
\\\hline\hline
 & \multicolumn{2}{c}{Expt.}    & DFT-LDA & DFT-GGA & \multicolumn{2}{c}{BKS} 
\\
                  & Ref.$^g$ & Ref.$^h$ & Ref.$^i$ & Ref.$^i$ & Ref.$^j$ & this work \\
\hline
\\[-2mm]
$a_0$ (\AA)        & -      & 4.914   & 4.8701& 5.0284 & 4.941 & 4.942  \\
$c_0$ (\AA)        & -      & 5.405   & 5.3626& 5.5120 & 5.405 & 5.448   \\
$\CC_{11}$ (GPa)     & 85.9   & 86.6    & 76.6  & 87.1  & 90.5  & 90.59  \\
$\CC_{12}$ (GPa)     & 7.16   & 6.74    & 5.88  & -7.82 & 8.1   & 8.06   \\
$\CC_{13}$ (GPa)     & 10.94  & 12.40   & 6.58  & 6.30  & 15.2  & 15.21  \\
$\CC_{14}$ (GPa)     & -17.66 & -17.80  & -17.8 & -17.0 & -17.6 & -17.68 \\
$\CC_{33}$ (GPa)     & 89.59  & 106.40  & 95.9  & 87.1  & 107.0 & 106.94 \\
$\CC_{44}$ (GPa)     & 57.77  & 58.0    & 54.1  & 49.1  & 50.2  & 50.23  \\
$\CC_{66}$ (GPa)     & 39.4   & -    & 35.3  & 47.5     & -     & 41.26  \\
\hline\hline
\end{tabular}
\newline
$^a$Reference~\citep{erhart_analytical_2005}, 
$^b$Reference~\citep{haynes_crc_2016},  
$^c$Reference~\citep{pastewka_screened_2013},  
$^d$Reference~\citep{karch_ab_1994}, 
$^e$Reference~\citep{balamane_comparative_1992}, 
$^f$Reference~\citep{kumagai_development_2007},
$^g$Reference~\citep{gregoryanz_high_pressure_elasticity_2000}, $^h$Reference~\citep{wang_elasticity_2015}, 
$^i$Reference~\citep{kermode_first_prinicple_force_field_2010},  $^j$Reference~\citep{vanBeest_BKS_1990}, 
$^k$Reference~\citep{nielsen_Si_1985},
$^l$Reference~\citep{pun_optimized_potential_silicon_2017},
$^m$Reference~\citep{lambrecht_calculated_SiC_1991},
$^n$Reference~\citep{tersoff_modeling_1989}
\end{table*}

For silicon, the relative error of the lattice constant and the elastic constants between our results and the reference publications is smaller than $0.01$\% for the \textsc{Kum} and the \textsc{TIII} potential.
For the \textsc{SW} potential we observe small relative errors in the order of $0.1$\% for the elastic constants, which include non-affine displacements, and a large relative error of $7.32$\% for $\cc_{44}$, which does not take non-affine displacements into account. 
This larger error for $\cc_{44}$ is probably related to the parameters of the finite difference approach in the reference publication \citep{balamane_comparative_1992}.  
For the \textsc{EAII} potential, we observe small relative errors of up to $2.4$\% in the elastic constants with respect to the reference values~\citep{erhart_analytical_2005}. 
This is likely due to the discrepancy in the lattice constant compared to the original publication. 
We repeated our computation using the reference lattice constant of $a=5.429~\text{\AA}$ and obtained $\CC_{11}=169.24~\text{GPa}$, $\CC_{12}=64.16~\text{GPa}$, $\CC_{44}=71.67~\text{GPa}$ and $\cc_{44}=110.99~\text{GPa}$.
Using the reference lattice constant decreases the maximal relative error of the elastic constants between our results and the reference values to $1.34$\%.  
Based on our results, the relative error between the elastic constants from experiments and DFT is smallest for the \textsc{Kum} potential.

For silicon carbide, we observe a maximal relative error between the reference values and our results of $0.46$\% for the \textsc{EA} potential and $3.8$\% for the \textsc{TIII} potential.
The largest error relative to experiments is $6.8$\% and $20.4$\%, while the error relative to the DFT values is at most $8.4$\% and $22.9$\% for the \textsc{EA} and the \textsc{TIII} potential, respectively.  
Although \textsc{EA} and \textsc{TIII} are based on the same functional form, the \textsc{EA} parametrization improves the accuracy of the lattice constant and the elastic constants with respect to the experimental and the DFT results significantly.

For $\alpha$-quartz the errors between our results and the original publication are less than $1.0$\%.
Despite the comparatively simple functional form of the \textsc{BKS} potential, the relative errors of the structural and elastic properties with respect to the reference data from experiments and DFT-LDA data are below $28$\% and $56$\%, respectively.
For the DFT-GGA data, the maximal relative error is about $200$\% which appears for the $\CC_{12}$ value.
Comparing experiments to the ab-initio computations, it is evident that the DFT-LDA results are in better agreement with the experimental values~\citep{kermode_first_prinicple_force_field_2010}. 
We further note that more complex potentials do not provide a significant improvement of the lattice constants and elastic properties of $\alpha$-quartz compared to the original \textsc{BKS} potential \citep{carre_new_fitting_2008,shan_charge_optimized_2010,kermode_first_prinicple_force_field_2010,lee_modified_SiO_2016}, but defects, interfaces, phase boundaries and other relevant property are significantly improved \citep{herzbach_comparison_2005,erhard_machine_learning_silica_2022}.

\subsection{Elastic constants at finite hydrostatic pressure}
\begin{figure*}[ht!]
	\includegraphics[]{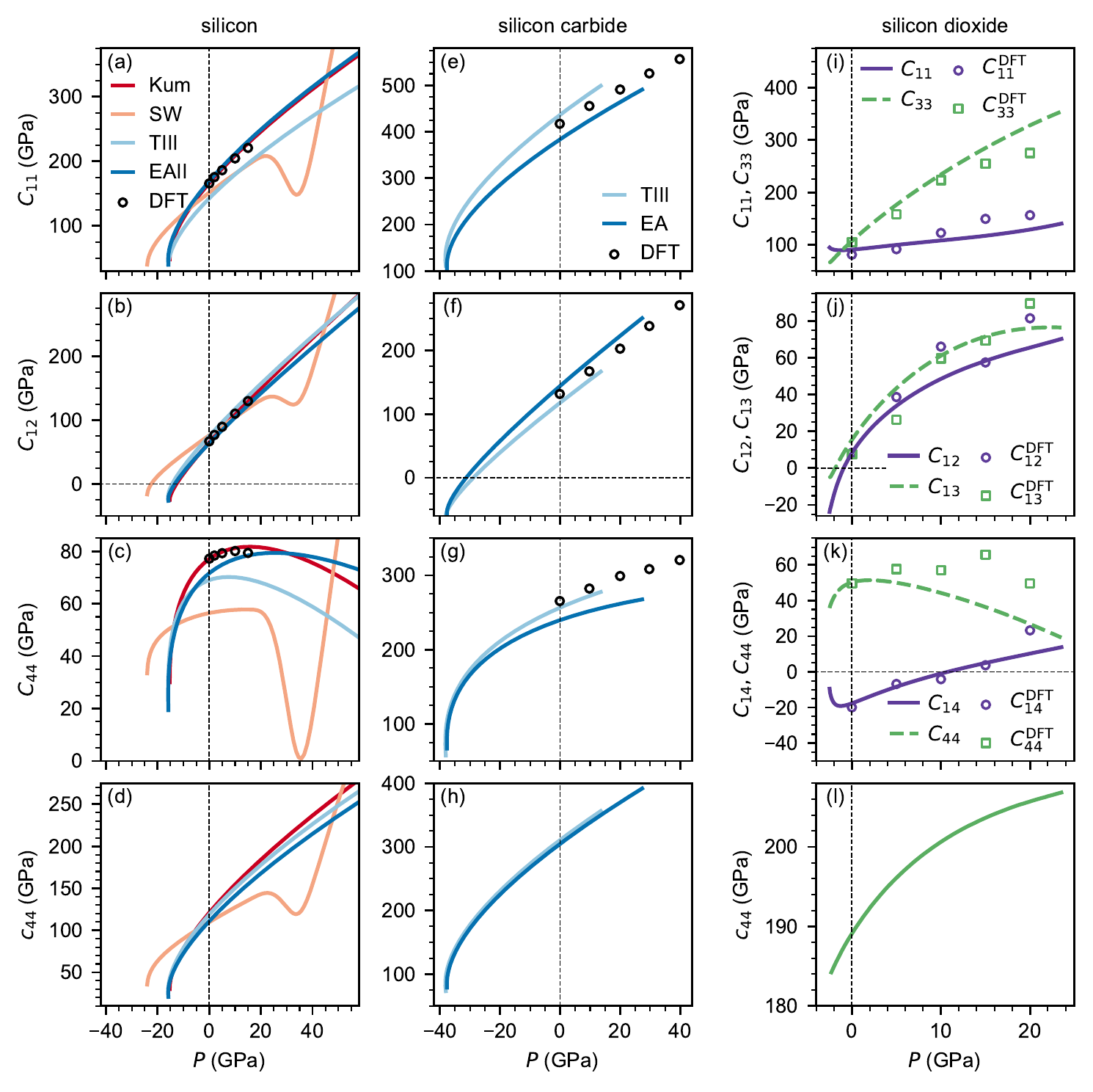}
 	\caption{\label{fig: stres_dep_elastic_moduli} Pressure-dependent elastic constants at zero temperature for silicon (a)-(d), silicon carbide (e)-(h) and silicon dioxide (i)-(l).
    DFT results for silicon are taken from \cite{karki_Si_DFT_1997}, for silicon carbide from \cite{lu_SiC_DFT_2008} and for silicon dioxide from \citep{kimizuka_SiO2_DFT_2007}.
    $\CC_{ij}$: elastic constants with non-affine contributions; $\cc_{ij}$: elastic constants without non-affine contributions.}
\end{figure*}

After validation of our approach at zero stress, we investigate the effect of hydrostatic pressure on the elastic properties.
In \cref{fig: stres_dep_elastic_moduli} we show the pressure-dependent elastic constants for all investigated materials and interatomic potentials. The vertical dashed lines mark zero hydrostatic pressure. 
For bond-order potentials under hydrostatic tension, we consider only the range of pressures where the first neighbor shell is within the first cutoff distance, corresponding to Si-Si bond distances of $r_{\text{Si,Si}}=2.7~\text{\AA}$ for \textsc{Kum} and \textsc{TIII} and $r_{\text{Si,Si}}=2.75~\text{\AA}$ for \textsc{EAII}.
For silicon carbide, this happens when the Si-C bond length exceeds $r_{\text{Si,C}}=2.2~\text{\AA}$.
Under compression, we consider the full range of pressures in order to highlight the impact of second-nearest neighbors on the elastic properties.
We first consider the elastic constants $\CC_{ij}$, which include the contribution from non-affine displacements. 
The qualitative behavior of the elastic constants of pure silicon is similar for all bond-order potentials.
Upon compression $\CC_{11}$ and $\CC_{12}$ increase monotonically while $\CC_{44}$ initially increases, reaches a maximum and subsequently decreases.
The location and the absolute value of that maximum varies slightly between the potentials.
Under tension all three elastic moduli decrease rapidly.
In contrast, the pressure dependence of the elastic constants for the \textsc{SW} potential is qualitatively different from the one observed for the bond-order potentials.
Under compression all moduli increase up to a pressure of $\approx 24~$GPa.
At this point, second-nearest neighbors enter the cutoff range and are included in the interaction.
Upon further compression the elastic moduli as a function of pressure drops towards a local minimum before starting to increase rapidly. 
Under tension the elastic moduli decrease, although the decrease is less rapid compared to the one seen for the bond-order potentials. 
We note that the effect of second-nearest neighbors on the elastic properties is a common problem for interatomic potentials with a finite interaction range~\cite{muser_interatomic_2022,muser_improved_cutoff_2022}.
Nevertheless, the large effect in \textsc{SW}-like potentials at comparatively small compressive pressures has not been reported before. 

In the case of silicon dioxide we present results only in the pressure range $ -3~\text{GPa} \leq P \leq 23~\text{GPa}$.
For smaller/larger pressures the initial lattice becomes unstable and would undergo a phase transformation (as discussed below in Sec.~\ref{sec:results_lattice_stability}) which results in a sudden change of the elastic constants.  
In the region where the initial crystalline structure of silicon dioxide is stable, the elastic constants, with the exception of $\CC_{44}$, increase with increasing compression.
In contrast, during tension all elastic moduli, except $\CC_{14}$, decrease.

In \cref{fig: stres_dep_elastic_moduli} (d), (h) and (l) we show the elastic modulus $\cc_{44}$ without the contribution from non-affine displacements. 
Due to symmetries in the lattice, the non-affine contribution in silicon and silicon carbide enters only in $\CC_{44}$, while for silicon dioxide the non-affine contribution is nonzero for all elastic moduli. 
If we neglect the non-affine contribution, the elastic modulus $\cc_{44}$ increases continuously under compression and decreases less rapidly under tension. 
This behavior is independent of the material and the interatomic potential, and is related to the stiffening of the effective pair interactions.
In direct comparison with the $\CC_{44}$ value, we observe that the non-affine part generally decreases the elastic constants.

\begin{figure}[]
    \centering
	\includegraphics[]{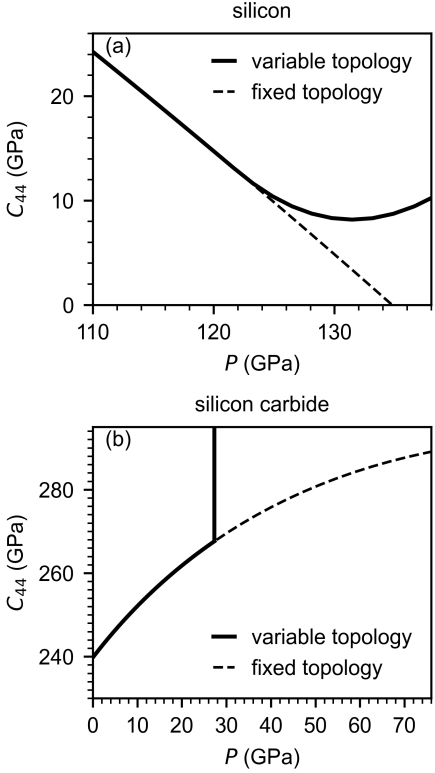}
	\caption{\label{fig:elastic_moduli_FT} 
    Effect of a fixed bonding topology versus a variable topology on the pressure dependence of $\CC_{44}$ for silicon~(\textsc{Kum}) (a) and silicon carbide~(\textsc{EA}) (b). The results deviate in regions where neighbors sit within the cutoff distance.
    }
\end{figure}

In the preceding paragraphs, we saw that the inclusion of second-nearest neighbors in some interatomic potentials affects the pressure dependent elastic constants in such a way that their behavior is no longer reliable.
In order to check if this is also true for the other potentials, e.g. \textsc{Kum}, we perform additional simulations using the fixed topology approach described in Sec.~\ref{sec: methods}.
In \cref{fig:elastic_moduli_FT} we show calculations of $\CC_{44}$ at fixed and variable topology.
The figure shows that at a compressive pressure of $\approx 123~$GPa, the elastic constant $\CC_{44}$ for silicon (\cref{fig:elastic_moduli_FT}(a)) shows increases rapidly if a variable topology is used.
(Note that we excluded those extreme pressures in the discussion above.)
If we rather use a fixed bonding topology in the \textsc{Kum} potential, this increase disappears and the $\CC_{44}$ curve continuously drops to zero, where the lattice looses stability (as discussed in the next section).
The reason for this sudden increase is once again the inclusion of second-nearest neighbors.
The pressure for this to occur is a factor of five larger than what is observed for the \textsc{SW} potential, and is related to the smaller cutoff radius $r_{c} ^{\textsc{Kum}}=3.0$~\AA ~$< r_{c}^{ \textsc{SW}}=3.77$~\AA. 
The inclusion of second nearest-neighbor therefore effectively stabilizes the crystal lattice and the system avoids the elastic instability.

\Cref{fig:elastic_moduli_FT}(b) shows once more the pressure-dependence of $\CC_{44}$ for silicon carbide modeled using the \textsc{EA} potential.
This time, we also show the effect of second nearest neighbors on the pressure-dependency of $\CC_{44}$.
At the pressure where second nearest neighbors enter the interaction range, $\CC_{44}$ increases instantaneously.
Using a fixed topology in the \textsc{EA} potential for silicon carbide does not change the pressure-dependency of $\CC_{44}$ and enables us to investigate the elastic constants for higher compressive pressures.
In the following section, which discusses lattice stability, we will only show results for potentials with fixed topology since the lattices typically mechanically collapse beyond the cutoff range.

\subsection{Stability of the crystal lattice under hydrostatic pressure \label{sec:results_lattice_stability}}
\begin{figure*}[ht!]
    \centering
	\includegraphics[]{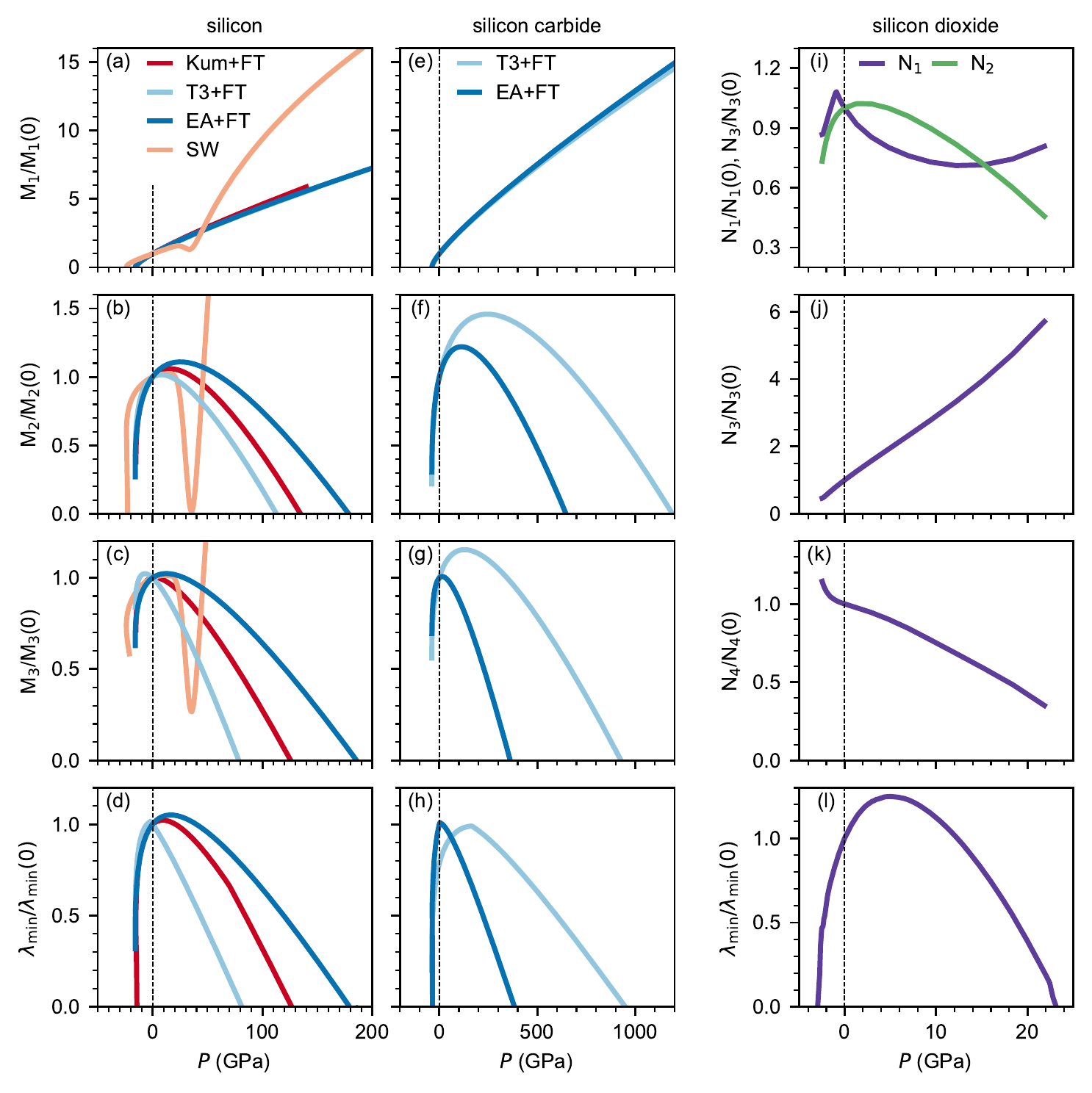}
	\caption{\label{fig: elastic_stability} Eigenvalues of the elastic tensor $\ff{C}$, denoted by $M_1$, $M_2$, $M_3$ for cubic solids and $N_1$, $N_2$, $N_3$, $N_4$ for silicon dioxide, and smallest eigenvalue $\lambda_\text{min}$ of the Hessian $\t{H}$ as a function of pressure $P$. Shown are results for silicon (a)-(d), silicon carbide (e)-(h), and silicon dioxide (i)-(l).
	The eigenvalues are normalized to their value at zero hydrostatic pressure.
 The systems loose mechanical stability when an eigenvalue becomes negative.
	}
\end{figure*}

So far, we have assumed the initial crystal structure to remain stable under arbitrarily large deformation. 
However, as already discussed in \cref{sec: lattice_stability}, instabilities related to the h-matrix or to atomic positions at fixed h-matrix can occur.
In the literature~\cite{wang_crystal_1993,mouhat_necessary_2014}, 
they are referred to elastic and dynamic instability, respectively.
We first focus on the elastic instability, which for cubic lattice symmetry reduces to the conditions~\citep{born_stability1_1940,wang_crystal_1993,mouhat_necessary_2014},
\begin{align}
\begin{split}
    M_1 &= \CC_{11} + 2\CC_{12} \geq 0, \\
    M_2 &= \CC_{44} \geq 0, \\
    M_3 &= \CC_{11}-\CC_{12} \geq 0,
    \label{eq: cubic_stability_condistions}
\end{split}
\end{align}
on the elastic constants for the lattice to be stable.
For silicon dioxide with trigonal lattice symmetry to become unstable, one of the following criteria needs to be violated \citep{mouhat_necessary_2014},
\begin{align}
\begin{split}
    N_1 &= \CC_{11} - \vert \CC_{12} \vert \geq 0,\\
    N_2 &= \CC_{44} \geq 0, \\
    N_3 &= \CC_{33}(\CC_{11} + \CC_{12}) - 2\CC_{13}^2 \geq 0, \\
    N_4 &= \CC_{44}(\CC_{11} - \CC_{12}) - 2\CC_{14}^2 \geq 0.
    \label{eq: rhombohedra_stability_condistions}
\end{split}
\end{align}
Here, $M_1$ to $M_3$ and $N_1$ to $N_4$ are the unique values of the eigenvalues of the tensor of elastic constants, some of which are degenerate.
In \cref{fig: elastic_stability}, we show these eigenvalues together with the smallest eigenvalue of the Hessian.

In the following, we refer to the pressure at which one of the eigenvalues $M_i$, $N_i$ or $\lambda_i$, vanishes as a critical pressure, $P^{\text{c}}$.
We first consider the behavior of pure silicon under hydrostatic compression using bond-order potentials. 
With increasing pressure, $M_1$ increases continuously while $M_2$ and $M_3$ decrease.
Thereby, $M_3$ is the vanishing eigenvalue for the \textsc{Kum+FT} and the \textsc{TIII+FT} potential.
$M_3$ vanishes at a critical pressure of $P^{\text{c}} \approx 125.7$~GPa and $P^{\text{c}} \approx 78.3$~GPa for the \textsc{Kum+FT} and the \textsc{TIII+FT} potential, respectively. 
Interestingly, for the \textsc{EAII+FT} potential $M_2$ vanishes first at a critical pressure of $P^{\text{c}} \approx 175.9$~GPa.
\textsc{TIII} and \textsc{EAII} have the same functional form, but their difference in critical pressure is significant.  
Using the \textsc{SW} potential, none of the stability criteria are violated under compression, but the eigenvalue $M_1$, $M_2$ and $M_3$ all have a minimum at a pressure $\approx 35.72$~GPa.
Under tension, $M_1$ vanishes for all bond-order potentials at an almost identical pressure of $P^{\text{c}} \approx -15$~GPa.
For the \textsc{SW} potential, $M_2$ vanishes at $P^{\text{c}} \approx -22.7$~GPa.
Comparing the critical pressures from an elastic instability with the ones from a dynamic instability, we observe that they occur at almost the same pressure.

For silicon carbide, which also has a cubic crystal structure, we observe that $M_3$ vanishes first in all interaction potentials.
The critical pressures are $P^{\text{c}} \approx 949.4$~GPa for \textsc{TIII+FT} and $P^{\text{c}} \approx 383.64$~GPa for \textsc{EA+FT}.
Under dilatation, $M_1$, vanishes first at approximately $P^{\text{c}}\approx-38$~GPa for both potentials.
We observe again that the critical pressure for the dynamic instability is in accordance with the one obtained from the elastic stability analysis.

For $\alpha$-quartz, we observe dynamic instabilities at critical pressures of $P^{\text{c}}\approx 23.3$~GPa and $P^{\text{c}}\approx-2.94$~GPa for compression and  tension.
As a result of these dynamic instabilities, the initial crystalline structure becomes unstable and transforms to a new stable crystalline state.
These phase transformations lead to an instantaneous change in the elastic constants for silicon dixoide and result in a sudden change of the elastic stability moduli $N_i$.
Since the elastic stability conditions are finite $N_i > 0$ at the critical pressure for the dynamic instability, we conclude that the dynamic instability is responsible for these phase transformation

\subsection{\label{sec: multiaxial_stability}Stability of the crystal lattice at multiaxial stress}

\begin{figure}[t!]
    \centering
	\includegraphics[]{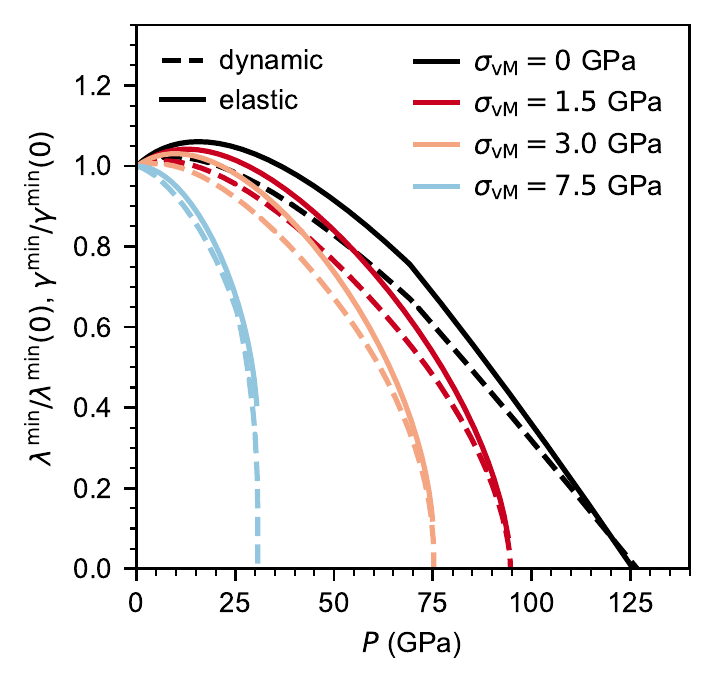}
	\caption{
    \label{fig: multiaxial} Smallest eigenvalue $\gamma^{\text{min}}$ of the elastic tensor $\ff{C}^\text{sym}$ and smallest eigenvalue $\lambda^{\text{min}}$ of the Hessian $\t{H}$ for diamond cubic silicon.
    }
\end{figure}

Deformation of solids in real world applications is rarely hydrostatic and instead complex multiaxial states of stress may arise, for example when contacting a surface with a sphere~\cite{Johnson1985-av}.
It is therefore important to consider the stability of a crystalline phase under such a deformation.
As an example of such a multiaxial deformation, we consider diamond cubic silicon subjected to a prescribed stress tensor 
\begin{equation}
    \tt{\sigma}
    =
    \begin{pmatrix}
      -P-\sigma_\text{vM}/3 & 0 & 0 \\
      0 & -P-\sigma_\text{vM}/3 & 0 \\
      0 & 0 & -P+2\sigma_\text{vM}/3 \\
    \end{pmatrix}
\end{equation}
where $\sigma_\text{vM}$ is the von Mises stress.
The interaction between atoms is modeled using the \textsc{Kum} potential with a fixed cutoff for $\sigma_\text{vM} \neq 0$ and the \textsc{Kum+FT} potential for $\sigma_\text{vM} = 0$.
To rule out the possible effects of the finite interaction range described above, we checked that the atoms interact only with nearest neighbors as long as the initial crystalline structure remains stable.
In contrast to the previous subsections, the deformation of the box is induced by prescribing the stress tensor $\tt{\sigma}$ and allowing the simulation cell to fluctuate.
The stability of the lattice is now determined by the symmetric part of the tensor of Birch coefficients $C_{ij}^{\text{sym}}$, which no longer has the original cubic symmetry because of the multiaxial stress state.
In \cref{fig: multiaxial}, we show the smallest eigenvalue $\gamma^{\text{min}}$ of $C_{ij}^\text{sym}$ together with the smallest eigenvalue of the Hessian $\lambda^{\text{min}}$ for certain values of $\sigma_\text{vM}$.
All eigenvalues have been normalized to their value at the first step of deformation, and we consider only the range of pressures until one of the eigenvalues, $\lambda^{\text{min}}$ or $\gamma^{\text{min}}$, becomes zero.
We observe that the values of $\gamma^{\text{min}}$ decrease with increasing compression, but always remain finite.
In contrast, the eigenvalues of the Hessian become zero at some critical pressure $P^{\text{c}}$, marking a phase transition induced by a dynamic instability.
Thereby, the critical pressure $P^{\text{c}}$ for the phase transition decreases with increasing $\sigma_\text{vM}$, i.e. with increasing deviation from the purely hydrostatic stress state.
Upon compressing the crystal above the critical pressure for the dynamic instability, the atoms instantaneously transform to a new stable crystalline structure, which here is the $\beta$-Sn (or Si-II) phase of silicon.
This change in simulation cell and atomic position is accompanied by sudden changes in the vibrational and elastic properties.

\section{Discussion}

This paper establishes a consistent framework for the calculation of finite stress elastic constants in molecular RVEs, including the effect of nonaffine atomic displacements.
It is important to reiterate that the elastic constants sensitively depend on the type of embedding medium.
The literature~\citep{leibfried_theory_1961,barron_second-order_1965,thurston_wave_1965,wallace_stability_1965,wallace_thermoelasticity_1967,hill_elasticity_and_stability_1975,hill_principles_1977,wang_crystal_1993,wang_mechanical_1995,wang_unifying_stability_2012,levitas_nonlinear_2021} typically discussed elastic constants at constant second Piola-Kirchhoff stress or at constant Cauchy stress.
The latter are also known as the Birch coefficients.
However, even these situations are highly idealized as they ignore stress gradients in the embedding medium that cannot be avoided unless this medium itself is perfectly matched, i.e. it consists of repeated images of the RVE.
This means, that mechanical stability of the RVE depends on the type of embedding.
Like most of the literature~\cite{wang_crystal_1993,mizushima_ideal_1994,wang_mechanical_1995,tang_atomistic_1995,karki_Si_DFT_1997,tang_fracture_1994,tang_atomic_size_effects_1995,choudhury_quartz_crystal_2006,kimizuka_SiO2_DFT_2007,lee_SiC_thermodynamic_2015,levitas_multiaxial_stress_2017}, we have restricted our example calculations to stability at constant Cauchy stress, but to what extend these calculation are representative of actual experimental conditions depends on the specific experiment.

Numerically, Hessians of RVEs are typically computed using finite differences.
We have here also derived analytical expressions for complex manybody potentials, that allow analytical computation of Hessians, vibrational modes, elastic constants and other linear repsonse properties.
This enables calculation of these properties for large heterogeneous systems, where finite differences become prohibitive or are impossible, such as near an instability.

As example calculations, we applied this method to a set of crystals.
All crystals under investigation show finite ranges of stability in experiments.
For silicon, diamond anvil cell experiments show a collapse of the diamond cubic lattice into $\beta$-tin at $12.7$~GPa~\cite{haberl_cSi_2013}.
This pressure roughly matches the equilibrium transition pressure obtained obtained from a tangent construction to the pressure volume curves of the respective phases in molecular calculations ($12.7$~GPa \textsc{TIII}~\cite{mizushima_ideal_1994} $9.8$~GPa \textsc{Kum}~\cite{Moras2018-lm}, $7.8$~GPa DFT-LDA~\cite{mizushima_ideal_1994}).
Our own calculations of the equilibrium diamond cubic to $\beta$-Sn transition show transition pressures between $11.8$ and $13$~GPa for the investigated interatomic potentials.
The ultimate limit of lattice stability occurs at much higher pressures ($125.7$~GPa \textsc{Kum}, $78.3$~GPa~\textsc{TIII}, $175.9$~GPa \textsc{EAII}).

For silicon-carbide, diamond anvil cells experiments with a pressure medium show an instability of 3C-SiC to a rocksalt structure at roughly $100$~GPa~\cite{yoshida_SiC_pressure_1993}.
DFT computations predict a slightly smaller critical pressure between $P^{\text{c}} \approx 58 ~$GPa and $P^{\text{c}} \approx 75~$GPa for this phase transition~\citep{lee_SiC_thermodynamic_2015,daviau_review_SiC_2018}.
Again, the ultimate stability of 3C-SiC from our calculations is much higher ($949.4$~GPa \textsc{TIII}, $383.4$~GPa \textsc{EA}).
Consistent with DFT calculations~\cite{lu_SiC_DFT_2008}, we find that a vanishing eigenvalue $M_3$ (tetragonal shear) leads to this instability. 
This is in contrast with results from Tang and Yip, who identified $M_2$ (simple shear) as the eigenvalue that vanishes first~\citep{tang_atomistic_1995,tang_atomic_size_effects_1995}.
In general, care has to be taken specifically for SiC, which has a large number of polymorphs.
Those polymorphs cannot be discriminated by the simple potentials used here that only consider nearest neighbor interactions.
In addition, SiC shows significant charge transfer.
A consideration of Coulombic interactions is likely necessary to capture the experimentally observed collapse to rocksalt, which is a prototypical ionic structure.

For silicon dioxide a crystal-to-crystal phase transition is observed in diamond anvil cell experiments at a pressure of roughly $21~$GPa~\cite{kingma_quartz2_1993}.
These experiments are supported by numerical computations which predict a transition from alpha quartz (quartz I) to quartz II phase.
Simulations predict the transition pressure to be between $16~$GPa and $22~$GPa, depending on the deviation from the purely hydrostatic stress state~\citep{watson_dynamical_stability_SiO2_1995,tse_high_pressure_phase_1997,campana_Quartz_phase_transition_2004,choudhury_quartz_crystal_2006}.
The experimental and computational estimated  critical pressure agrees well with the dynamical instability $P^\text{c}\approx 23.3~$GPa in our simulations.

The ultimate stability of crystalline lattices (as characterized by a dynamical or elastic instability) appears to reached in experiment on silicon dioxide, but does not appear to play a role in the stability of silicon and silicon carbide.
In particular for silicon, the experimentally observed Si-I to Si-II transition appears where the high pressure phase becomes thermodynamically stable and not where the lattice collapses.
Transition between the two states must then involve nucleation and growth processes, as transitioning directly between two crystalline phases involves barriers that scales with sample volume.

However, there are multiple factors that lower the ultimate limit of stability as obtained from our zero-temperature calculations.
First, temperature softens the elastic response and helps overcome energy barriers, which can significantly affect the critical pressure.
For example, \citeauthor{mizushima_ideal_1994} used finite temperature calculations to show that the the critical pressure of lattice stability in silicon reduces from $105$~GPa at zero temperature to $64$~GPa at room temperature~\citep{mizushima_ideal_1994}.
Second, it is difficult to achieve perfect hydrostatic conditions in diamond anvil cell experiments.
For example, anvil cell experiments using ethanol-methanol as a pressure medium have been reported to deviate from a purely hydrostatic condition at around $10$~GPa pressure ~\cite{angel_DAC_hydrostatic_2007,takemura_hydrostaticity_DAC_2021}.
Our calculations on silicon show that multiaxial stress significantly reduces the limit of lattice stability.
As shown in \cref{fig: multiaxial}, a shear stress of magnitude equal to $10$\% of the hydrostatic pressure reduces the critical pressure roughly by a factor of two.
A combination of temperature fluctuations and multiaxiality therefore likely lowers the limit of lattice stability significantly, potentially to the pressure of the purely thermodynamic transition.

\section{Summary \& Conclusions}
In this paper, we revisited the question of how to define elastic constants at an arbitrary state of stress and included the role of non-affine displacements.
Based on these theoretical results, we gave closed-form expressions for a select set of many-body interatomic potentials for all terms required to analytically compute the elastic constants in atomistic simulations.
These terms include expressions for the Hessian, the non-affine forces and the Born elastic constants for generic many-body potentials.
These analytical expressions are implemented in our open-source software \textsc{matscipy}~\citep{matscipy} and they have the advantage of being exact.
Their derivation is based on a generalized, unified functional form that fits many empirical interatomic potentials.
It is easily extendable to interatomic potentials beyond those presented here.
The resulting elastic properties are fast to compute and not prone to parameters in the numerical computation.

We demonstrated these methods on the elastic constants and the lattice stability of silicon, silicon carbide and silicon dioxide under volumetric and multiaxial deformation.
We highlighted that all employed bond-order potentials and cluster potentials using a finite interaction range suffered from unreliable results once second-nearest neighbors are included in the interaction range.
Furthermore, we showed that multiaxiality plays a critical role in the stability limits of crystalline lattices.

\begin{acknowledgements}
We thank Patrick Dondl and Wolfram G. Nöhring for useful discussion. The authors acknowledge support from the Deutsche Forschungsgemeinschaft (DFG, grants PA 2023/2 and 461911253 ``AWEARNESS''). All molecular simulations are carried out using \textsc{ASE}~\citep{hjorth_larsen_atomic_2017} and \textsc{matscipy}~\citep{matscipy}. Simulations were carried out on NEMO at the University of Freiburg (DFG grant INST 39/963-1 FUGG).
\end{acknowledgements}

% Retore the \v macro to its original meaning
\let\v\vczech
%\bibliography{literature}


\end{document}